\input{aipcheck}

\documentclass[
    ,final            
  ]
  {aipproc}

\layoutstyle{8x11single}
\newcommand{\be}{\begin{equation}}
\newcommand{\ee}{\end{equation}}
\newcommand{\beq}{\begin{eqnarray}}
\newcommand{\eeq}{\end{eqnarray}}

\begin{document}

\newcommand{\twopt}[5]{\langle G_{#1}^{#2}(#3;\mathbf{#4};\Gamma_{#5})\rangle}
\newcommand{\threept}[7]{\langle G_{#1}^{#2}(#3,#4;\mathbf{#5},\mathbf{#6};\Gamma_{#7})\rangle}

\title{Hadron Deformation and Form Factors from Lattice QCD}


\author{C. Alexandrou}{
  address={ Department of Physics, University of Cyprus, CY-1678
Nicosia, Cyprus}
}



\begin{abstract}
We review the current status of lattice QCD studies of the nucleon system.
In particular, we focus  on the determination of 
the shape of the nucleon by probing its wave function as well
as by evaluating the $N$ to $\Delta$ transition form factors. 
\end{abstract}

\maketitle


\section{I. Introduction}
Lattice QCD provides an 
indispensable method in our efforts to
solve Quantum Chromodynamics (QCD).
It is the only approach, using directly the QCD Lagrangian,
available up to now
in the energy regime between
 high energy processes, where
perturbation theory is applicable, and very low energies, where
chiral perturbation theory can be used.
Lattice QCD is 
a discretized version of QCD
 formulated in terms of path integrals on a space-time lattice~\cite{Wilson}
 with only
parameters the bare quark masses and the coupling constant, like the
continuum theory. One recovers continuum physics  by 
extrapolating results obtained at finite lattice spacing $a$ to $a=0$.
In order to perform the continuum extrapolation a separate calculation at
several  
values of $a$ is required. 
Because this can require large computer resources, 
in most cases, the
 strategy is
to work close to the continuum limit, either by choosing $a$ sufficiently
small or by improving the discretization scheme to eliminate order-$a$ 
terms.   
Numerical evaluation of observables in lattice QCD
necessarily requires that the
 size of the lattice is finite.  To keep the
size of the box large enough to fit the hadrons well inside one must
increase the number of sites as one decreases $a$. Therefore
calculations at decreasing values of $a$ require increasingly larger
computer resources. To keep finite volume effects small
one must have a box that is much larger than the Compton wavelength
of the pion. Usually we take $Lm_\pi\stackrel{>}{\sim} 5$ where
$L$ is the spatial length of the box and $m_\pi$ the pion mass.
In an actual calculation the bare quark masses for the $u$ and $d$ quarks
are taken larger than in the real world. Given that  computational costs
increase like $m_\pi^{-9}$, the use of larger quark masses enables
inversion of the fermionic matrix, needed for the calculation of
hadronic matrix elements, with
currently available  resources. To extract
physical quantities from 
lattice calculations typically done with pion masses above $350$~MeV,
one needs to extrapolate to the physical quark masses. 
It is only very recently that we can
reliably reach  pion masses,
 below $350$~MeV~\cite{MILC,tmQCD}.
This is a very important step forward in our effort to  eliminate
one source of systematic error associated with the extrapolation to 
the light quark masses. 
Typical state-of-the art lattices use $a\stackrel{<}{\sim}0.1$~fm and
$L\sim 3$~fm for pions of mass down to about 350~MeV.

Discretization of space-time  introduces an ultra-violet
cut-off limiting the highest momentum to $2\pi/a$. Therefore, the lattice
provides a regularization of the ultra-violet divergences of the theory
making it  well defined from the start.
 Working in a
 finite box allows
only discrete values of momentum in units of $2\pi/L$.
The bare coupling constant and quark masses are
tuned as $a$ changes to leave physical quantities unchanged.
Besides using a finite lattice spacing, a finite
volume and heavier u- and d- quarks, an additional
 step, that enables  us to numerically evaluate the path integrals
needed,
is rotation to imaginary time, $t\rightarrow -it$, resulting in replacing
 $\exp(iS)$ by   $\exp(-S_E)$
where $S$ ($S_E$) is the classical QCD action in Minkowski (Euclidean)
space.
Having a real action one can apply stochastic
 methods commonly used in statistical mechanics to
evaluate the path integrals.
Whereas finite $a$, volume and quark masses are amenable to
systematic improvements, rotation to Euclidean space 
selects a set of observables that can be  studied within this framework.
These are observables that can be determined from the properties of
the discrete lower lying states. For this set of observables, lattice
QCD produces the exact answer provided the extrapolations to the continuum 
and  infinite volume limits as well as 
to the physical quark masses are under control.

In a typical lattice calculation one starts
  by choosing the bare coupling constant  $g$,  which fixes the lattice
spacing, and the bare masses for the u-, d- and s-quarks. One
then computes a physical quantity such as the mass of the pion and the nucleon
in lattice units as a function of the  quark mass. The
pion mass is used to fix the u- and d-  quark masses (assumed degenerate)
 and the mass of the kaon or $\phi$ to fix the strange quark mass whereas the
lattice spacing is determined by extrapolating the results,
for instance, for the nucleon mass to the physical pion mass. 
Any other physical quantity
in the light quark sector then follows.
If instead of the nucleon mass one chooses
 another physical quantity such as the pion decay constant,
$f_\pi$, to set $a$ the resulting value should be the same  if finite lattice
volume and non-zero $a$ effects are under control.

Like in the continuum, the lattice 
QCD action, $S_E=S_g[U]+S_F[U,\bar{\psi},\psi]$, has a purely gluonic part
$S_g$ written in terms of  the gauge link
 $U_\mu(n)=\exp(iagA_\mu(n))$, which connects
site $n$ with $n+1$ in the $\mu$-direction and a fermionic part $S_F$, which
contains the kinetic energy of the quarks and the interaction terms.
$A_\mu(n)$ denotes the gluon field.
Gauge symmetry is exactly preserved by $S_E$. 
As in any renormalizable theory, the fermionic action is bilinear 
in the fermion
fields and can be written in terms of the fermionic matrix
$D$ as $S_F=\sum_{n,j}\bar{\psi}(n)D_{nj}\psi(j)$. The exact form of
$D$ depends on the discretization scheme used for the fermions. The simplest is
due to Wilson~\cite{Wilson}. It has been widely studied but has the 
disadvantage of breaking explicitly chiral symmetry. Recent theoretical 
developments made it possible to have chiral fermions on the 
lattice. There are two equivalent formulations known as domain 
wall fermions~\cite{kaplan,shamir}
and overlap fermions~\cite{Hasenfratz,Neuberger}. 
They both require larger computer resources
than Wilson fermions. 

The vacuum expectation value of any gauge invariant
operator $\hat{O}$ can be computed by evaluating the path integral
\be
<\Omega|\hat{O}|\Omega> =\frac{\int d[U]d[\bar {\psi}]d[\psi]\>\> 
O[U,\bar{\psi},\psi] 
e^{-S_g[U]-S_F[U,\bar{\psi},\psi]} }{\int d[U]d[\bar {\psi}]d[\psi] 
e^{-S_g[U]-S_F[U,\bar{\psi},\psi]} }\quad.
\label{expectation value}
\ee
Integrating over the fermionic degrees of freedom we obtain
\be
<\Omega|\hat{O}|\Omega> =\frac{\int d[U] \>\>\det(D[U])O[U,D^{-1}[U]] 
e^{-S_g[U]}}{Z} \quad\quad Z\equiv \int d[U]\> \>\det(D[U]) e^{-S_g[U]} \quad, 
\label{expectation value2}
\ee
where a factor $D^{-1}_{jn}[U]$ substitutes each appearance of
 $-\bar{\psi}_n \psi_j$ in $O$.
One can now perform the path integrals numerically by stochastically 
generating a representative ensemble
of gauge fields $U$ according to the probability 
$\exp\left \{-S_g[U]+\ln\left(\det(D[U])\right)\right \}/Z$ and then compute
\be
 <\Omega|\hat{O}|\Omega>=\lim_{N\rightarrow \infty}\frac{1}{N}\sum_{k=1}^N O[U^k,D^{-1}[U^k]]\quad,
\label{numeric value}
\ee
  which involves the evaluation of the inverse of the fermionic
matrix. For a typical lattice of size $24^3\times 48$ the dimension of the
complex matrix $D$ is $8$ million by $8$ million. 
Therefore the time consuming part
of a lattice calculation is the generation of an ensemble of
 gauge configurations
and the  computation of the inverse of the fermionic matrix $D$ which
yields the quark propagator. In many applications only a column of 
$D^{-1}$ is required.

The first lattice calculations
were performed in the quenched approximation, which neglects
pair creation by setting $\det(D)=1$ in 
Eq.~\eqref{expectation value2}. This facilitates the generation of gauge
links since one is left with the local action $S_g[U]$. 
In the quenched approximation 
the program outline above, i.e.
taking the continuum and infinite volume limits and extrapolating to the
physical quark masses, has been carried out for the spectrum
of the low lying hadrons~\cite{QCDPAX}. 
During the past five years theoretical progress in combination with
terascale computers have made unquenched calculations with light pions and
large enough volumes feasible~\cite{MILC,ETMC,newSESAM,DESY}, 
using a number of   different discretization schemes.

In order to study the role
of the pion cloud, which is expected to provide
an important ingredient in the description of the
 properties of the nucleon system,
one must generate dynamical 
gauge configurations with light quarks using large volumes.
 In this
work  the light quark regime is  studied in two ways:
\begin{enumerate}
 \item We use  configurations with the lightest available
dynamical Wilson fermions.
The unquenched
configurations are simulated with two degenerate flavors of 
Wilson fermions~\cite{newSESAM,DESY}. 
\item We use
MILC configurations generated with two degenerate light and one 
strange staggered
quarks  using the Asqtad improved action~\cite{MILC}. For the
valence quarks we use domain wall fermions 
 that  preserve chiral
symmetry on the lattice. This is therefore a hybrid calculation that
 uses different fermions for the sea and valence quarks,
i.e the  matrix $D$ appearing in the determinant in Eq.~\eqref{expectation value2}
is different from $D^{-1}$ involved in the calculation of $O$ in Eq.~\eqref{numeric value}. Such hybrid calculations
have been
successful in recent evaluations of fundamental physical quantities
such as $g_A$~\cite{gaxial} and the pseudoscalar decay constants $f_\pi$
and $f_K$~\cite{orginos}.
\end{enumerate}
Bearing in mind that  both quenched and unquenched Wilson fermions
have discretization errors of order  $a$,
and that both Asqtad and domain wall fermion actions have discretization
 errors of order $a^2$ and fermions preserving chirality, in contrast to
Wilson fermions, agreement between the results
within these two
 different lattice fermion formulations provides a non-trivial
 check of consistency of the lattice results.
The hybrid calculation is computationally the most demanding
 since it requires propagators on a five-dimensional lattice.
The bare quark mass for the domain wall fermions, the size of
the fifth dimension and the renormalization
factors $Z_V$ and $Z_A$ for the four-dimensional vector and axial 
vector currents are
 taken from Ref.~\cite{gaxial}.
The parameters of our calculations are given in Table~\ref{table:parameters}.

\begin{table}[h]
\caption{
 We give the number of configurations,
 the hopping parameter, $\kappa$, which determines the bare quark mass 
via the relation $2am_q=(1/\kappa-1/\kappa_c)$ for the
case of  Wilson fermions or the mass of the u and d  quarks, $m_l$,
 for the case of staggered quarks,
the pion,  nucleon and $\Delta$ mass in lattice
units with their statistical errors determined from a jackknife analysis.
For Wilson fermions $a$
is set using the nucleon mass at the chiral limit
whereas for staggered fermions we take the value
extracted from the  static $q\bar{q}$ force
as determined in Ref.~\cite{MILCa}.} 
\label{table:parameters}
\begin{tabular}{|c|c|c|c|c|}
\hline
\multicolumn{1}{|c|}{no. of configurations } &
\multicolumn{1}{ c|}{$\kappa$ (Wilson) or $am_l$ (staggered) } &
\multicolumn{1}{ c|}{$a\>m_\pi$ } &
\multicolumn{1}{ c|}{$a\>M_N$ } &
\multicolumn{1}{ c|}{$a\>M_{\Delta}$ } 
\\
\hline
\multicolumn{5}{|c|}{Quenched $32^3\times 64$ \hspace*{0.5cm} $a^{-1}=2.14(6)$ GeV}
 \\ \hline
  200            &  0.1554 &  0.263(2) & 0.592(5)   & 0.687(7) \\
  200            &  0.1558 & 0.229(2) &  0.556(6)   & 0.666(8) \\
  200            &  0.1562 & 0.192(2) &  0.518(6)   & 0.646(9)\\
    &  $\kappa_c=$0.1571   & 0.       &  0.439(4)  & 0.598(6)\\
\hline
\multicolumn{5}{|c|}{Unquenched Wilson $24^3\times 40 $  \hspace*{0.5cm}
$a^{-1}=2.56(10)$ GeV} 
 \\\hline
 185                &  0.1575  & 0.270(3) & 0.580(7) & 0.645(5)\\
 157                &  0.1580  & 0.199(3) & 0.500(10) & 0.581(14) \\
\hline
\multicolumn{5}{|c|}{Unquenched Wilson $24^3\times 32 $  \hspace*{0.5cm}
$a^{-1}=2.56(10)$ GeV} 
\\\hline
 200                &  0.15825 & 0.150(3) & 0.423(7)  & 0.533(8)  \\
                    & $\kappa_c=0.1585$& 0. & 0.366(13)& 0.486(14)\\
\hline
\multicolumn{5}{|c|}{MILC $20^3\times 64 $  \hspace*{0.5cm}
$a^{-1}=1.58$ GeV} 
 \\\hline
 150                &  0.03  & 0.373(3) & 0.886(7) & 1.057(14)\\
 150                &  0.02  & 0.306(3) & 0.800(10)&  0.992(16)\\
\hline
\multicolumn{5}{|c|}{MILC $28^3\times 64 $  \hspace*{0.5cm}
$a^{-1}=1.58$ GeV} 
\\\hline
 118                &  0.01 & 0.230(3) & 0.751(7)  & 0.988(26)  \\
\hline
\end{tabular}
\end{table}

The main goal of this program is to calculate within lattice QCD the 
fundamental physical quantities of the nucleon-$\Delta$ system.
Providing the complete set of form factors and coupling constants
constitutes a very important input for model builders 
and for fixing the parameters of chiral effective theories.  In
this presentation we discuss
our results for the nucleon elastic form factors and the electromagnetic
and axial $N$ to $\Delta$ transition form factors as well as 
show first results on their wave functions. 
Other recent lattice studies on
nucleon properties can be found in
Ref.~\cite{others}.

\section{II. Lattice techniques}
The vacuum expectation 
value of gauge invariant operators  is computed by numerical evaluation
of appropriately defined path integrals. 
Let us first consider the
 evaluation of hadron masses. The vacuum expectation
value of the time ordered product
$G^h(t,{\bf q})=<\Omega|\sum_{\bf x}
\exp\left(i{\bf q}.{\bf x}\right)\hat{T} \hat{J}_h({\bf x},t) \hat{J}_h^\dagger(0)|\Omega>$ 
can be evaluated
using Eqs.~\eqref{expectation value2} and \eqref{numeric value} which require
inverting $D$ for each non-degenerate quark flavor
once per gauge configuration. The interpolating
fields, $\hat{J}_h(x)$ are operators in the Heisenberg representation
that create a trial state with the
quantum numbers, $h$, of the hadron that we want to study.
For example in the nucleon case an appropriate interpolating field is 
$\hat{J}_N(x)=\epsilon_{abc}\left[u_a{^T}(x)C\gamma_5d_b(x)\right]u_c(x)$ where 
$C$ is the charge conjugation operator and Latin indices denote
color quantum numbers. The large
time behavior of the correlator $G^h(t,{\bf 0})$ yields the 
mass:
\beq
G^h(t,{\bf q})&=&\sum_{n,{\bf p}}\>\>\sum_{\bf x}e^{-i{\bf q}.{\bf x}}
<\Omega|
e^{\hat{H}t}e^{-i\hat{{\bf p}}.{\bf x}} \hat{J}_h
e^{-\hat{H}t}e^{i\hat{{\bf p}}.{\bf x}}|n,{\bf p}><n,{\bf p}|\hat{J}_h^\dagger|\Omega>=
\sum_n |<\Omega|\hat{J}_h|n,{\bf q}>|^2 e^{-E_n({\bf q}) t} \nonumber \\
 G^h(t,{\bf 0}) &\>&\stackrel{t(m_{h_1}-m_{h_0})>>1}{\Longrightarrow}
|<\Omega|\hat{J}_h|h_0,{\bf 0}>|^2\> e^{-m_{h_0} t}
\label{hadron mass}    
\eeq
where $E_n({\bf q})=\sqrt{m_n^2+{\bf q}^2}$,
$|h_0>$ is the lowest eigenstate of QCD with  quantum numbers $h$ 
 with mass $m_{h_0}$ and we
have taken ${\bf q}={\bf 0}$. For large
time separation $t$  between the source and the sink the unknown
overlap factor $|<\Omega|\hat{J}_h|h_0,{\bf 0}>|^2$ and exponential
dependence cancel in the ratio
$m_{\rm eff}(t)=-\log\left[G^h(t)/G^h(t-1)\right]$, 
which therefore becomes
time independent  and can be fitted to a constant to extract 
the mass, $m_{h_0}$, of the lowest state.

Three-point functions of the form
$G^{h A \tilde{h}}(t,t_1;{\bf q})=<\Omega|\sum_{\bf x,y}\
e^{i{\bf q}.{\bf x}}\hat{T} \hat{J}_{h}({\bf y},t)\hat{A}({\bf x},t_1) 
\hat{J}_{\tilde{h}}^\dagger(0)|\Omega>$ are required for the evaluation 
of form factors. 
Inserting a complete set of hadronic states between operators in 
$G^{h A \tilde{h}}(t,t_1;{\bf q})$ as we
 did above for the extraction of hadron masses  from two-point
functions, we obtain
\beq
G^{h A \tilde{h}}(t,t_1;{\bf q})&=&\sum_{k,n,{\bf p}^\prime,{\bf p}}\sum_{\bf x,y}\>\>e^{i{\bf q}.{\bf x}}
<\Omega|\hat{J}_{h}|n,{\bf p}^\prime> e^{-E_n({\bf p}^\prime) t} 
e^{i{\bf p}^\prime.{\bf y}}
e^{E_n({\bf p}^\prime) t_1} e^{-i{\bf p}^\prime.{\bf x}}
<n,{\bf p}^\prime|\hat{A}|k,{\bf p}>
e^{-E_n({\bf p}) t_1} e^{i{\bf p}.{\bf x}}
<k,{\bf p}|\hat{J}_{\tilde{h}}^\dagger|\Omega>
\nonumber \\
&\>& \stackrel{\Delta E (t-t_1)>>1,\tilde{\Delta E} t_1>>1}{\Longrightarrow}
<\Omega|\hat{J}_h|h_0,{\bf 0}><\tilde{h}_0,{\bf p}|\hat{J}_{\tilde{h}}^\dagger|\Omega>
<h_0,{\bf 0}|\hat{A}|\tilde{h}_0,{\bf p}> e^{-m_{h_0}(t-t_1)} e^{-E_{\tilde{h}_0}({\bf p})t_1}
\label{form factors}
\eeq
where $\Delta E$ and $\tilde{\Delta E}$
 are the energy differences between the two lowest 
 hadronic states with quantum numbers $h$ and $\tilde{h}$
 and ${\bf p}=-{\bf q}$.
The exponential time  dependence 
and unknown overlaps can be canceled by dividing with appropriate
combinations of two-point functions $G^h(t,{\bf q})$. For example 
in the ratio
$
R=G^{h A \tilde{h}}(t,t_1;{\bf q})/\sqrt{G^h(2t-2t_1,{\bf 0})G^{\tilde{h}}(2t_1,{\bf q})}
$
the overlap factors and exponentials cancel in the large time limit when 
the ground states $h_0$ and $\tilde{h}_0$ dominate,
yielding the matrix element $ <h_0,{\bf 0}|\hat{A}|\tilde{h}_0,{\bf p}>$.
Since both two- and three- point functions decay 
exponentially it is crucial in constructing the ratio to choose 
combinations of two-point functions
that involve the shortest possible time separations and to
use techniques that isolate the lowest hadronic states $|h_0>$ and 
$\tilde{h}_0>$ at short time intervals. The former can be done
by choosing better but more complicated ratios than the one given here, 
 to be discussed in Sections IV and V~\cite{PRL,NN},
 whereas the latter by constructing better interpolating
 fields. 
Smearing techniques are routinely used
for achieving   ground state dominance
before the signal from the time correlators
is lost in the noisy large time limit.
We use  gauge invariant Wuppertal smearing to replace local by smeared
quark operators
 at the source and the sink~\cite{Guesken-Alexandrou}
and, when needed, we apply
 hypercubic averaging~\cite{HYP} of the gauge links that enter 
the construction of the Wuppertal smearing function.
As can be seen from Eq.~\eqref{form factors}, 
to compute three-point functions two  sums over the spatial
volume are needed. Performing the Wick contractions on the quark level 
for the three-point function one
finds expressions that involve  the full inverse of the fermionic
matrix $D$ (all-to-all propagator). This is to be contrasted
with two-point functions for which only one spatial
sum enters
and therefore
only one column of $D$ is needed.
 The technique
to automatically  do one of the spatial sums by using
 an appropriately defined  input vector 
that uses one column of the inverse of $D$ when inverting,
is known as sequential inversion. One can choose which of the two
sums to do first. In this work we do the sum over the sink ${\bf y}$.
This requires that  the quantum numbers of the sink 
are fixed but allows 
 any operator $\hat{A}$
with any momentum ${\bf q}$ to be inserted. 
This means that with one sequential
inversion one can extract the matrix element 
 $ <h_0,{\bf 0}|\hat{A}|\tilde{h}_0,{\bf p}>$
for given hadronic states but different operators and momentum ${\bf q}$.
We therefore  measure the matrix element for 
 all lattice momentum vectors that 
result in a given momentum transfer squared $q^2$, thereby
obtaining many statistically independent evaluations of the same
form factors reducing statistical noise.

\section{III. Probing hadron wave functions}
 
\begin{figure}[h]
\begin{minipage}{5cm}
\hspace*{-3.5cm}
\includegraphics[height=.2\textheight]{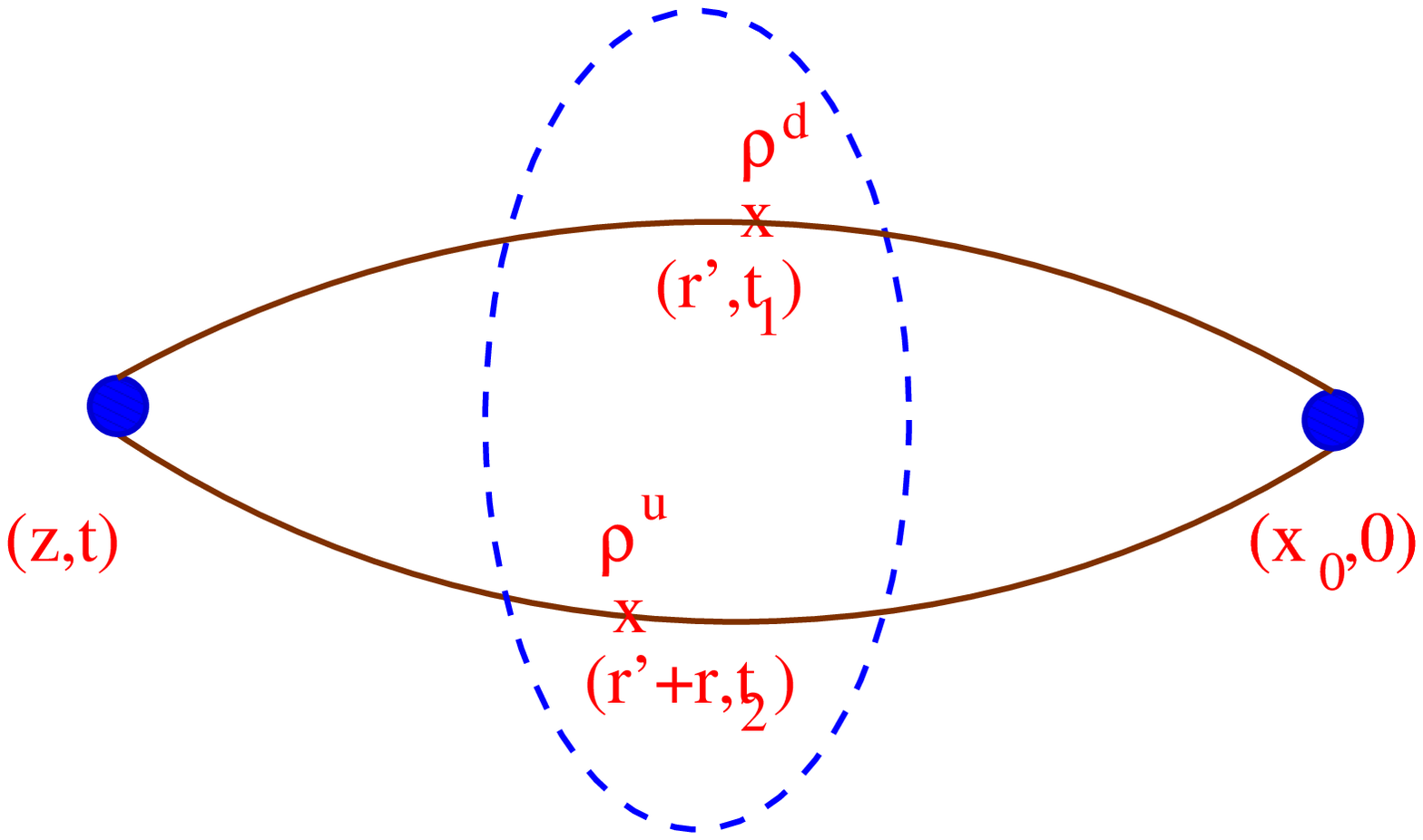}
\end{minipage}
\begin{minipage}{5cm}
\beq
 D({\bf r},t_1,t_2) &=& \int\> d^3r'\>
\langle h|\rho^{u}({\bf r}'+{\bf r},t_2)\rho^{d}({\bf r}',t_1)|h\rangle
\label{2density} \nonumber \\ 
\rho^f({\bf r},t) &=& :\bar{f}({\bf r},t)\gamma_0 f({\bf r},t):
\eeq
\parbox{8cm}{where for the density operator we use the normal order}
\parbox{8cm}{product so that disconnected graphs are excluded.}
\end{minipage}
\caption{The density-density correlator, $D({\bf r},t_1,t_2)$, for a meson. 
The time separations
$t_1$, $t_2$,
 $t-t_1$ and $t-t_2$
are taken large enough to isolate the mesonic ground state.}
\label{fig:2density}
\end{figure}

 Our main motivation for studying density-density correlators is that
 they reduce, in the non-relativistic limit,
 to the wave function squared yielding
detailed, gauge-invariant information on the internal structure 
of hadrons~\cite{Negele2,Wilcox}. 
The shape of hadrons, which is the topic of this workshop,
is one such important quantity that can be directly studied.
There are indications from experimental measurements of the
quadrupole strength in the  $\gamma^*N \> \rightarrow \> \Delta$ 
transition~\cite{cnp} that the nucleon and/or the $\Delta$ are deformed.
Density-density correlators,
 shown schematically in Fig.~\ref{fig:2density}, are four-point functions
since they involve the insertion of two density operators.
Four-point functions
are technically harder to compute than three-point functions 
discussed in the previous 
Section.
In particular, they require computation of all the spatial
columns of the inverse of the fermionic matrix requiring $L^3$ inversions.
 A straight forward computation of such an inverse is therefore
prohibitively expensive. In our first study~\cite{AFT}, 
which we consider as a feasibility
study, the density-density correlators were evaluated without explicit 
projection to zero momentum hadronic states. This means that higher
momentum states are suppressed only by the Euclidean time evolution.

\begin{figure}
\begin{minipage}{3.5cm}
\hspace*{-5cm}
 \includegraphics[height=.3\textheight]{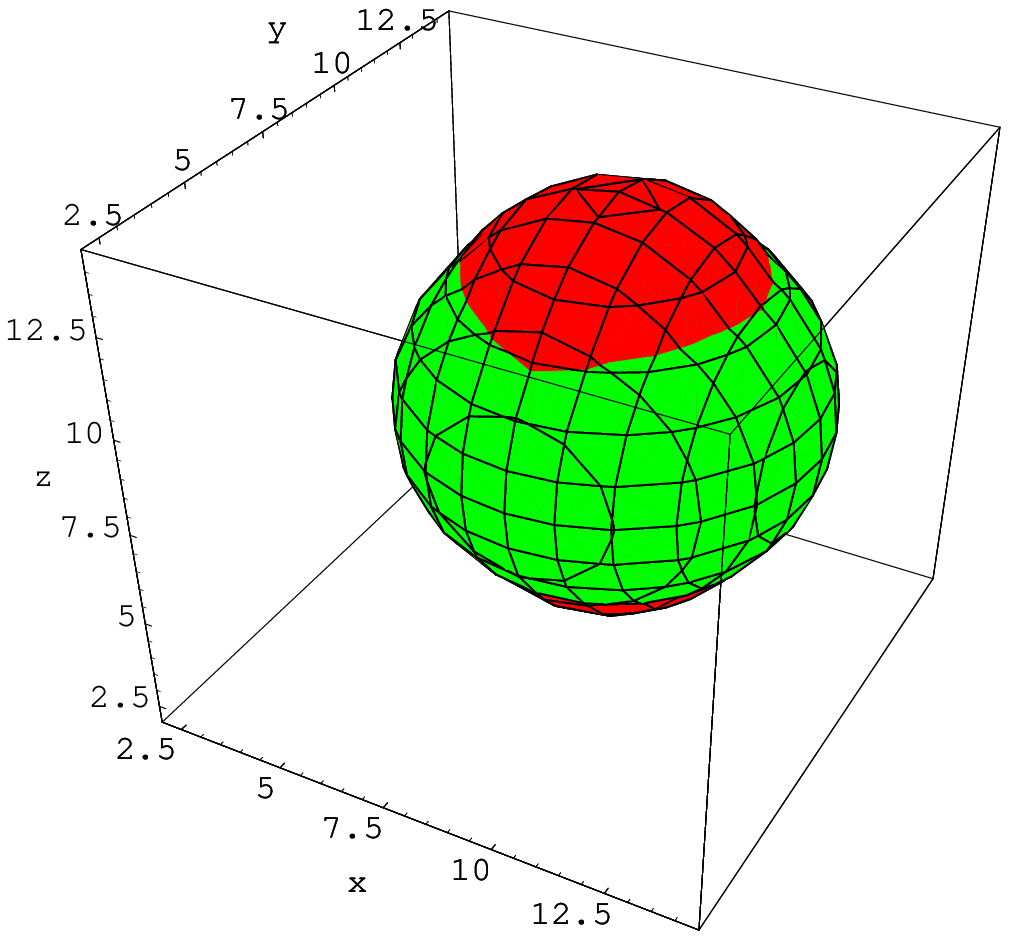}
\end{minipage}
\begin{minipage}{3.5cm}
 \includegraphics[height=.3\textheight]{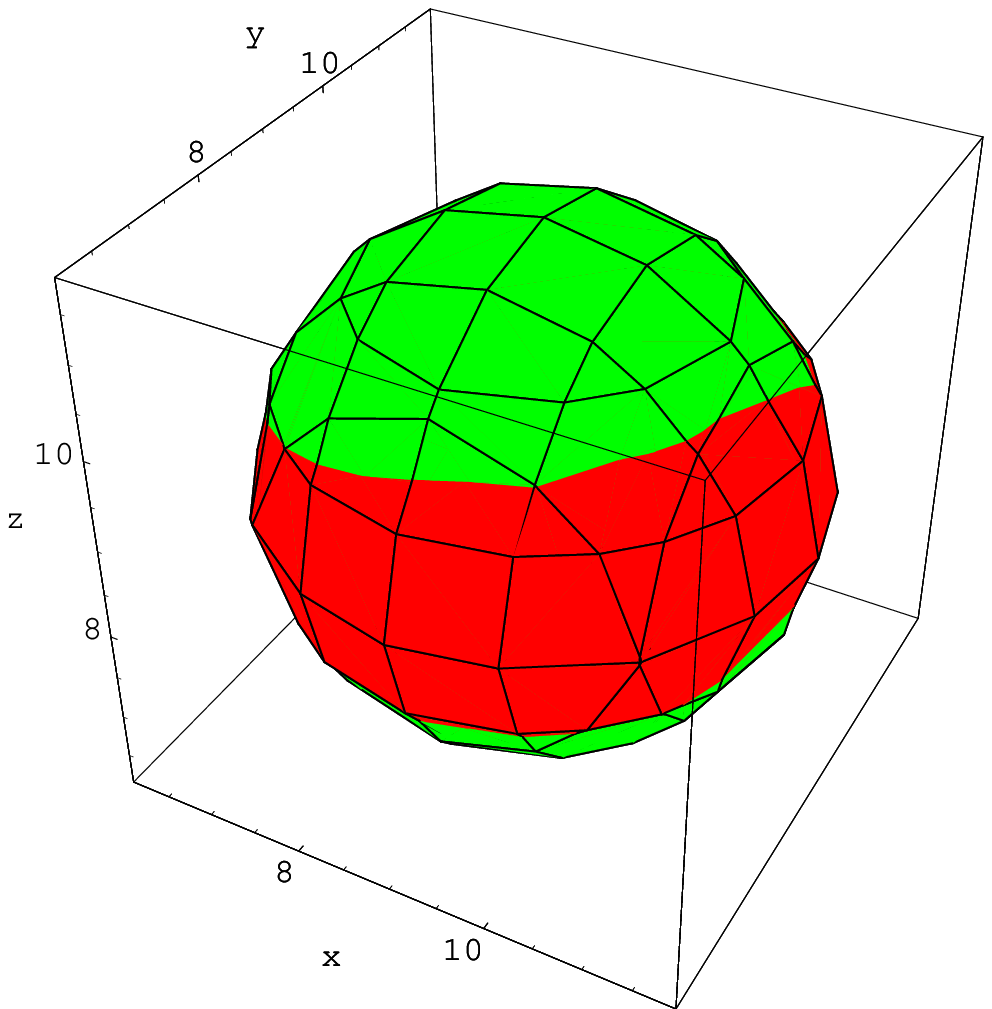}
\end{minipage}
\vspace*{0.5cm}
\caption{A three-dimensional contour plot of the correlator is shown in red.
Left: for the rho state with 0 spin projection (cigar shape);
  Right: for the $\Delta^+$ state with
+3/2 (slightly oblate) spin projection
for two dynamical quarks at $\kappa=0.156$.
Values of the correlator (0.5 for the rho, 0.8 for the $\Delta^+$) were chosen
to show large distances but avoid finite-size effects.
We have included for comparison
 the contour of a sphere (green).}
\label{fig:3D-contour}
\end{figure}

We have analyzed 220  quenched configurations at $\beta=6.0$
for   a lattice of size $16^3 \times 32$ obtained
from the NERSC archive, using the Wilson Dirac operator with
hopping parameter $\kappa=0.15, 0.153, 0.154$ and $ 0.155$.
 Using the relation
$2am_q=(1/\kappa-1/\kappa_c)$, with the critical value $\kappa_c=0.1571$,
one can obtain the naive quark mass $m_q$. A physical dimensionless 
quantity that 
is sensitive to the value of the bare quark mass is 
the ratio of the pion mass to the rho mass, which
 at these values of $\kappa$,  is
$0.88,\>0.84,\>0.78$ and 0.70 respectively.
  We  fix the source and the sink for maximum time separation,
which for this lattice 
is  $t/a=16$
 given the antiperiodic  boundary conditions in the 
time direction. 
Both 
density insertions are taken at the same time slice at the middle of the
time interval. To investigate the
importance of dynamical quarks, we
use  SESAM configurations~\cite{SESAM} generated 
with two dynamical degenerate
quark species at $\beta=5.6$ on a lattice of size $16^3\times 32$
 at $\kappa=0.156$ and 
 $0.157$.
The ratio  of the pion mass to rho mass is 0.83 at $\kappa=0.156$ and
0.76 at $\kappa=0.157$. These values are close to the quenched mass ratios
measured at $\kappa=0.153$  and $\kappa=0.154$  respectively,
allowing us to make pairwise quenched-unquenched comparisons.
The general conclusion of this comparison is that, at these heavy quark
masses, unquenching effects are small. 
We show in Fig.~\ref{fig:3D-contour}
contour plots of the density-density correlator for  the rho and the $\Delta$
for the case of dynamical Wilson fermions at
the heavy quark mass. The elongation
in the rho is clearly visible whereas the $\Delta^+$ appears to be
squeezed.

We have recently developed techniques for the  evaluation
of density-density  correlators with explicit projection to zero momentum 
hadronic states~\cite{latt05} and 
 we are in the process of  analyzing unquenched SESAM configurations
at lighter quark masses as listed in Table~\ref{table:parameters}.
  
\section{IV. Nucleon form factors}
 Recent results from polarization experiments~\cite{Jeff, Jager}, 
  have shown a qualitative different behavior  than the 
traditional Rosenbluth separation for the ratio
of the proton electric to magnetic form factor, $\mu_p G_E^p/G_M^p$.
An accurate determination of these
form factors in lattice QCD can provide
an important  theoretical input for understanding  the  $q^2$-dependence 
of this ratio. 
 We present here  results obtained using Wilson fermions
on the quenched and unquenched lattices given in Table~\ref{table:parameters}.
The smallest $q^2$-values are accessible on the  quenched lattice, 
which has the largest
spatial extent, 
since the smallest available non-zero
momentum on a finite
 lattice is $2\pi/L$ giving  $-q^2\sim 0.17$~GeV$^2$. 
 Although large momentum 
transfers are in principle available on typical  lattices,  
the Fourier transform of two- and three-point functions 
becomes noise-dominated for momentum transfers
beyond about 2~GeV$^2$, limiting the range of high $q^2$ values that
can be extracted accurately.

The standard decomposition of
 the nucleon electromagnetic matrix element 
for real or
virtual photons is given by
\beq
 \langle \; N (p',s') \; | j_\mu | \; N (p,s) \rangle &=&  
 \biggl(\frac{ M_N^2}{E_{N}({\bf p}^\prime)\;E_N({\bf p})}\biggr)^{1/2} 
  \bar{u} (p',s')  \biggl[\gamma_\mu F_1(q^2) 
+  \frac{i\sigma_{\mu\nu}q^\nu}{2M_N} F_2(q^2)\biggr] u(p,s) \; ,
\label{NjN}
\eeq
where $p(s)$ and $p'(s')$ denote initial and final momenta (spins), 
$ M_N$ is the nucleon mass,
 $F_1(0)=1$ for the proton since we have a conserved current and
$F_2(0)$ measures the anomalous magnetic moment. They are connected to the  
electric, $G_E$, and magnetic, $G_M$, Sachs form factors by the relations
\be
G_E(q^2)= F_1(q^2) + \frac{q^2}{(2M_N)^2} F_2(q^2) \hspace*{0.5cm}
G_M(q^2)= F_1(q^2) + F_2(q^2) \quad .
\label{Sachs ff}
\ee
The electromagnetic matrix element is extracted 
from the three-point function
$G^{Nj_\mu N}$ 
following the procedure outline  in the Section~II.
We use the lattice conserved   electromagnetic current,
\be
j^\mu (x)=\sum_{f=u,d}q_f\biggl[ \bar{\psi}^f(x+\hat{\mu})(1+\gamma_\mu) U^\dagger_\mu(x)\psi^f(x)-
\bar{\psi}^f(x)(1-\gamma_\mu) U_\mu(x)\psi^f(x+\mu)\biggr]
\label{conserved current}
\ee
symmetrized on site $x$ by taking
$
j^\mu (x) \rightarrow \left[ j^\mu (x) + j^\mu (x - \hat \mu) \right]/ 2
$.
We look for a plateau in  
the large Euclidean
time behavior of the improved ratio 
\beq
R (t, t_1;  {\bf q}\; ; \Gamma ; \mu) &=&
\frac{\langle G^{Nj_\mu N}(t, t_1 ;  {\bf q};\Gamma ) \rangle \;}{\langle G^{N}(t, {\bf 0};\Gamma_4 ) \rangle \;} \> \nonumber 
\biggl [ \frac{ \langle G^{N}(t_2-t_1, {\bf p};\Gamma_4 ) \rangle \;\langle 
G^{N} (t_1, {\bf 0};\Gamma_4 ) \rangle \;\langle 
G^{N} (t, {\bf 0};\Gamma_4 ) \rangle \;}
{\langle G^{N} (t_2-t_1, {\bf 0};\Gamma_4 ) \rangle \;\langle 
G^{N} (t_1, {\bf p};\Gamma_4 ) \rangle \;\langle 
G^{N} (t, {\bf p};\Gamma_4 ) \rangle \;} \biggr ]^{1/2} \nonumber \\
&\;&\hspace*{-1cm}\stackrel{\Delta E(t-t_1) \gg 1, \Delta E t_1 \gg 1}{\Rightarrow}
\Pi( {\bf q}\; ; \Gamma ; \mu) \; .
\label{R-ratio}
\eeq
where we show explicitly the dependence
on  the projection matrices 
$
\Gamma_j = \frac{1}{2}
\left(\begin{array}{cc}\sigma_j & 0 \\ 0 & 0 \end{array}
\right),\;j=1,2,3$ and
$\Gamma_4 = \frac{1}{2}
\left(\begin{array}{cc} I & 0 \\ 0 & 0 \end{array}
\right) \;\;
$
for the Dirac indices.
$\Delta E$ is the energy difference between the nucleon
and its excited $P_{11}$-state
and ${\bf q}=-{\bf p}$ for a  nucleon at rest in the 
final  state.
Since we want to study the
$q^2$-dependence of the form factors we evaluate the three point
function with sequential inversion through the sink.
We fix the source-sink time separation $t/a=11 (12)$ 
for the quenched (unquenched) Wilson
lattices and search for a plateau of 
 $ R(t,t_1; {\bf q}\; ; \Gamma ;\mu)$ as a function of
 the time slice, $t_1$,  at which  $j_\mu$ couples to a quark. 
$Q^2=-q^2$ denotes the Euclidean momentum transfer squared. 

\begin{figure}[h]
\begin{minipage}{3.5cm}
\hspace*{-5cm}
 \includegraphics[height=.3\textheight,width=2.35\textwidth]{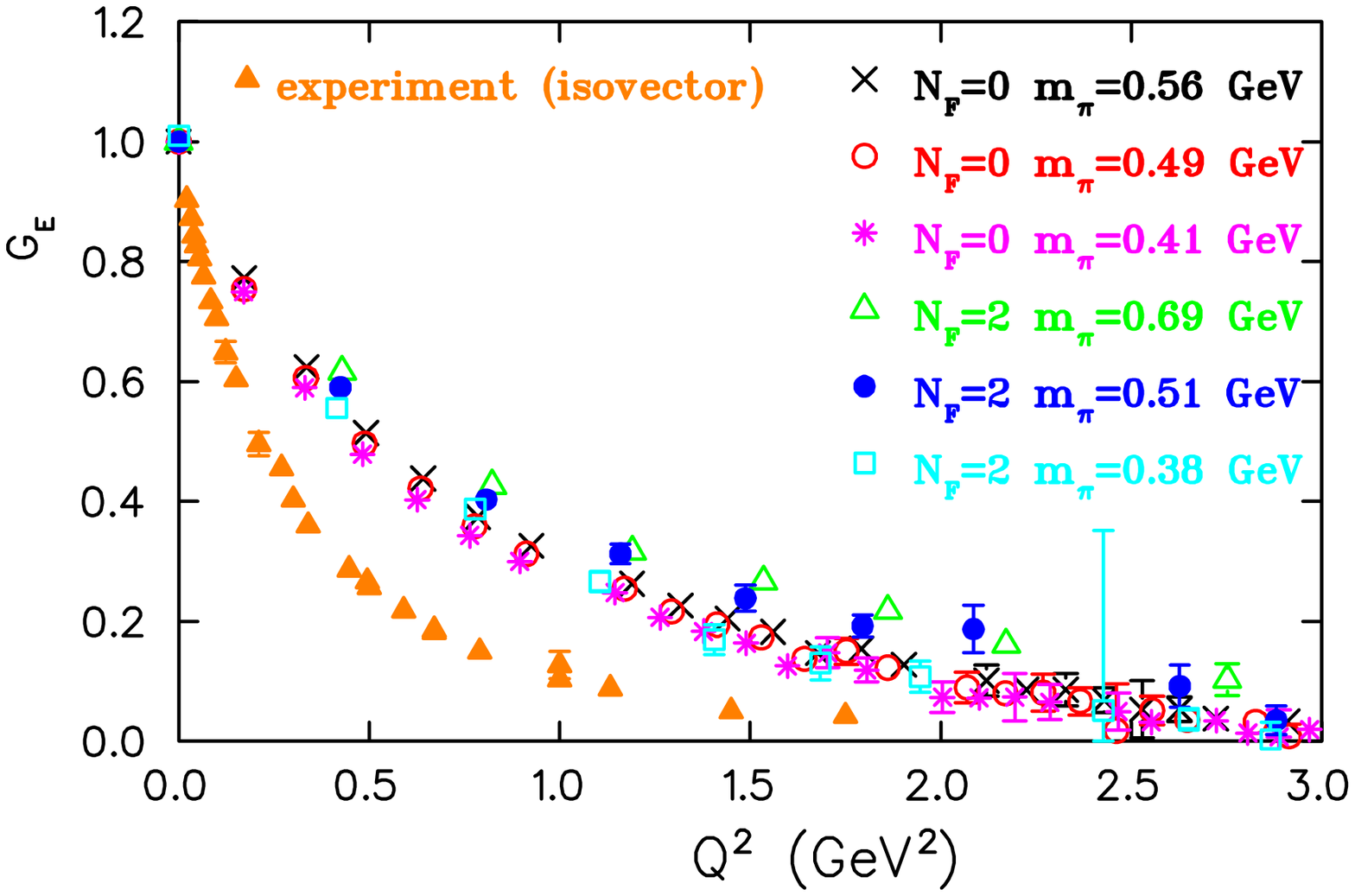}
\end{minipage}
\begin{minipage}{3.5cm}
\includegraphics[height=.3\textheight,width=2.35\textwidth]{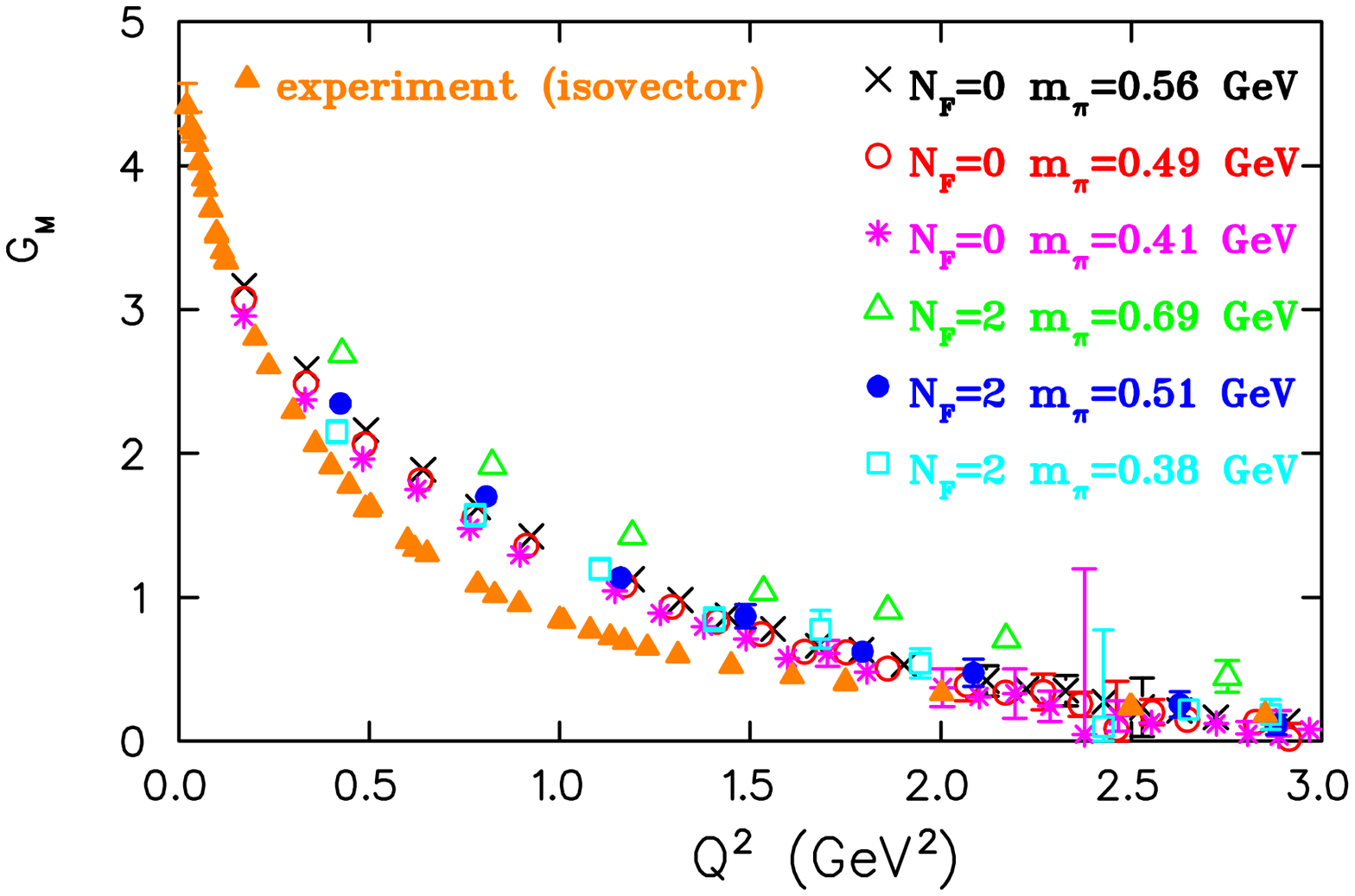}
\end{minipage}
\caption{The isovector form factors, $G_E$, (right) and 
 $G_M$, (left) as a function of $Q^2$. 
 By $N_F=0$ we denote quenched lattice results  at
$\kappa=0.1554$ (crosses), at $\kappa=0.1558$ (open circles) 
and at $\kappa=0.1562$ (asterisks). 
Results using two degenerate flavors of dynamical Wilson fermions
 are denoted by $N_F=2$ at $\kappa=0.1575$ 
(open triangles),
$\kappa=0.1580$ (filled circles)
and at $\kappa=0.15825$ (open squares).
The filled triangles show  
 experimental results for the isovector form factors
extracted by interpolating the experimental data for the
proton and neutron form factors (the details
of the interpolations)are given in Ref.~\cite{NN}).
}
\label{fig:GE and GM}
\end{figure}

The two Sachs form factors are extracted from the ratio defined in
Eq.~(\ref{R-ratio}) by choosing
appropriate combinations of the direction $\mu$ of the electromagnetic current
  and projection matrices $\Gamma$.
Provided the Euclidean time separations $t_1$ and $t - t_1$ are large
enough to filter the nucleon ground state, the ratio 
$R (t, t_1;  {\bf q}\; ; \Gamma ; \mu)$
becomes time independent and
the two form factors can be extracted from 
 the   Euclidean space expressions,    
\be
 \sum_{k=1}^3\Pi ({\bf
q}\; ;
\Gamma_k ;\mu=i) =  \frac{K}{2M_N} \biggl\{ (p_2-p_3)\delta_{1,i}
 + (p_3-p_1)\delta_{2,i} + (p_1-p_2)\delta_{3,i} \biggr\}
G_M(Q^2) 
\label{GM optimal}
\ee
\be \Pi ( {\bf q}\; ; \Gamma_4 \; ; \mu=i )  = K
\frac{q_i}{2 M_N} \; G_E (Q^2) \quad, \quad  
 \Pi ( {\bf q}\; ; \Gamma_4 \; ; \mu = 4)  = K
\frac{E_N +M_N}{2 M_N} \; G_E (Q^2)
\label{GE}
\ee
%
where $K=
\sqrt{\frac{2 M_N^2}{E_N(E_N + M_N)}}$ is a factor due to the
normalization of the lattice states. 
We note that the expression for $G_M$ is obtained by using
an optimal linear
combination for the nucleon sink that provides the maximal set of
lattice measurements from which  $G_M$
can be extracted requiring only one sequential inversion.
Eqs.~(\ref{GE})
 yield $G_E$ with an additional sequential inversion.

The $\gamma^* N\rightarrow N$ transition, in addition to 
an isovector part, contains 
isoscalar photon
contributions. This means that disconnected loop diagrams also
contribute. These are generally difficult to
evaluate accurately since the all-to-all quark
propagator is required. In order to avoid
disconnected diagrams, we calculate the isovector form factors.
 Assuming $SU(2)$ isospin
symmetry, it follows that 
\be 
\langle \; p \,| (
\frac{2}{3}\bar{u} \gamma^{\mu}u - \frac{1}{3}\bar{d} \gamma^{\mu}
d ) | p \rangle  - \langle \; n | ( \frac{2}{3}\bar{u}
\gamma^{\mu}u - \frac{1}{3}\bar{d} \gamma^{\mu} d ) | n \rangle \;
 = \langle \; p \, | ( \bar{u}
\gamma^{\mu} u - \bar{d} \gamma^{\mu} d ) | p \rangle . 
\ee
One can therefore calculate directly the three-point
functions related to the right hand side of the above relation, from which
the {\it isovector } nucleon form
factors 
\be
G_E (q^2) = G^p_E (q^2)\, - G^n_E
(q^2) , \hspace*{0.5cm}
 G_M (q^2) = G^p_M(q^2)- G^n_M (q^2) \quad,
\label{isovector}
 \ee
can be extracted using Eqs.~\eqref{GM optimal}
and \eqref{GE} by only evaluating   connected
diagrams.

Besides using an optimal nucleon source, the other
important ingredient in the extraction of the form factors,
 is to take into
account simultaneously
in our analysis  all the lattice momentum vectors that contribute to a given 
$Q^2$. This is done by solving the overcomplete set of equations
$
P({\bf q};\mu)= A({\bf q};\mu)\cdot F(Q^2) 
$
where $P({\bf q};\mu)$ are the lattice measurements of the ratio
given in Eq.~(\ref{R-ratio}) having statistical errors
$w_k$ and using the different sink types,
$F =  \left(\begin{array}{c}  G_{E} \\
                                    G_M \end{array}\right)$
and $A$ is an $M\times 2$ matrix which depends on 
kinematical factors with $M$ being the number of current 
directions and momentum vectors contributing to 
a given $Q^2$. We extract the form factors by 
minimizing 
\be
\chi^2=\sum_{k=1}^{N} \Biggl(\frac{\sum_{j=1}^2 A_{kj}F_j-P_k}{w_k}\Biggr)^2
\ee
using the singular value decomposition of the matrix $A$.
Given the fact
that one can have a few hundred lattice momentum vectors contributing 
for a given $Q^2$,  the statistical precision is highly improved. 
There is an additional advantage arising from including momentum vectors
${\bf q}$ as well as $-{\bf q}$ in our analysis:
 The lattice conserved current given in Eq.~\eqref{conserved current}
 differs from the local electromagnetic 
current $\bar{\psi}(x) \gamma_\mu \psi(x)$ by terms of order $a$. However
when we average over ${\bf q}$ and $-{\bf q}$ these   order $a$ terms
vanish.

\begin{figure}[h]
\begin{minipage}{3.5cm}
\hspace*{-5cm}
\includegraphics[height=.5\textheight,width=2.35\textwidth]{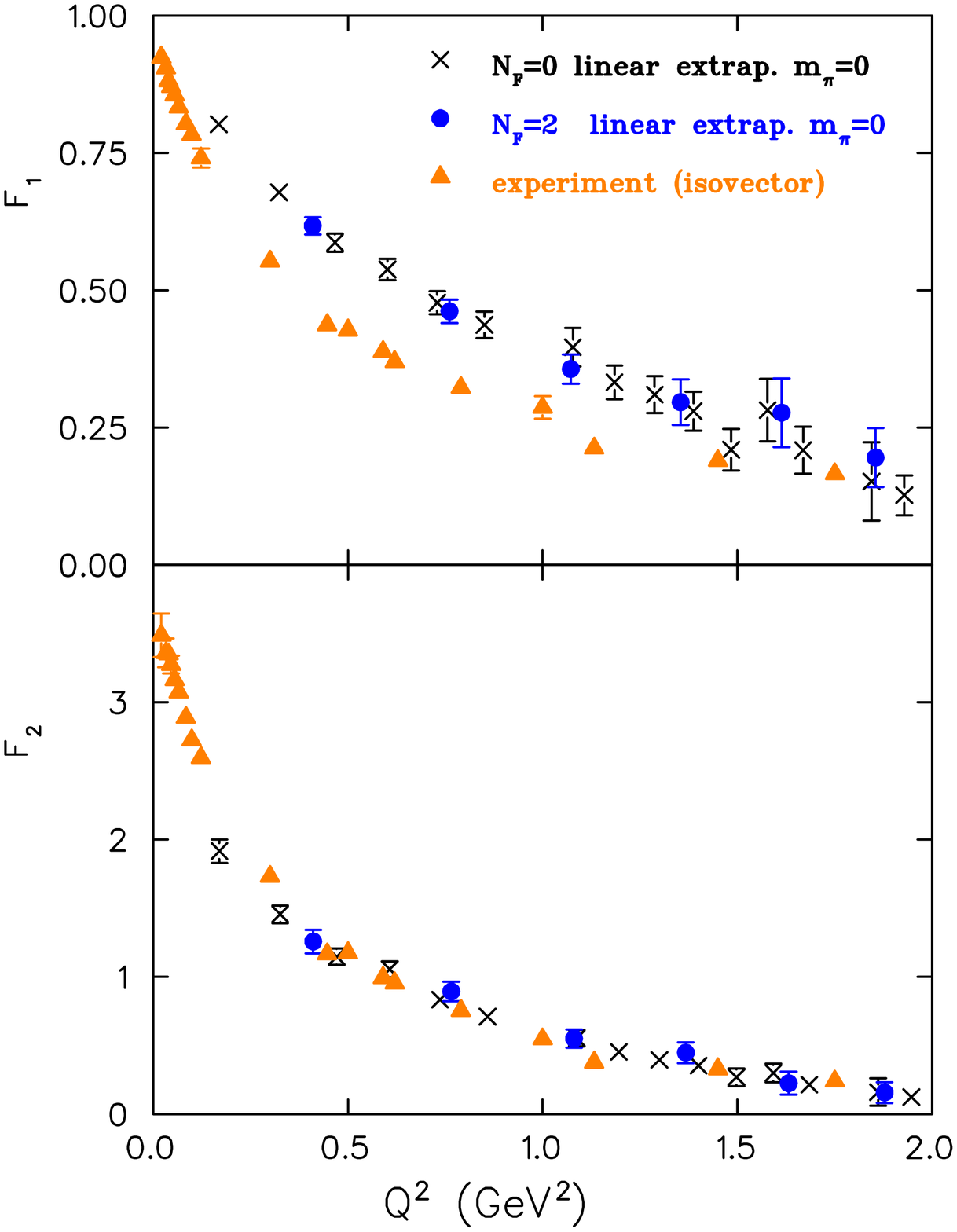}
\end{minipage}
\begin{minipage}{3.5cm}
\includegraphics[height=.5\textheight,width=2.35\textwidth]{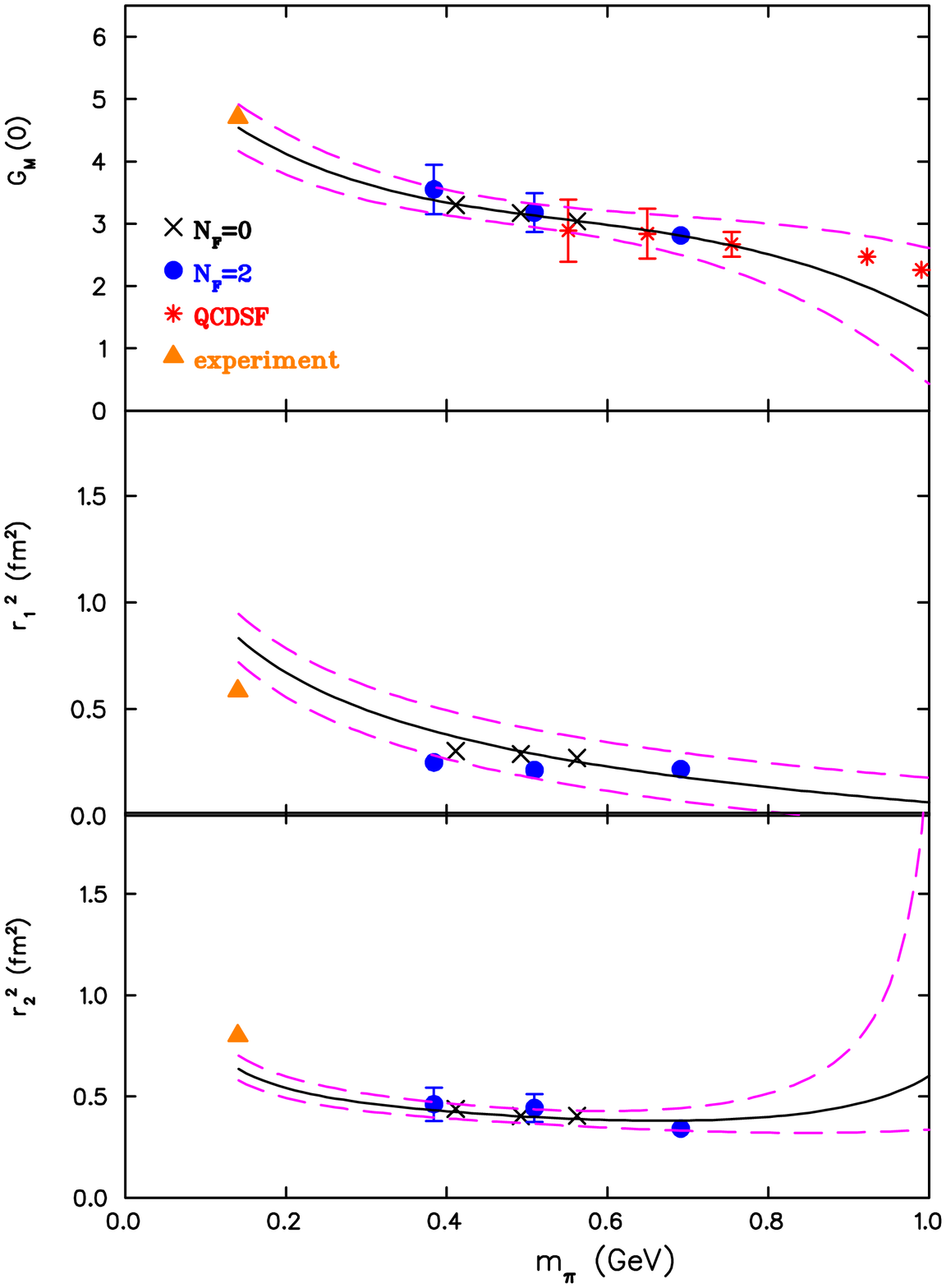}
\end{minipage}
\vspace*{-0.5cm}
\caption{Left: The form factors $F_1$  (upper) and $F_2$ (lower)
 as a function of  $Q^2$ at the chiral limit.
  The results
  extracted from experiment are shown by the filled
triangles.
Right: Chiral extrapolation of the magnetic moment (upper)  
and the r.m.s radii  $r_1$ (middle) and $r_2$ (lower). The solid line is 
the best fit to the effective chiral theory results. The dashed lines
show the maximal allowed error band using the errors on the fitted
parameters. In all graphs  quenched results are shown by the crosses
and  unquenched results by the filled circles. Quenched results 
from Ref.~\cite{QCDSF} are shown by the asterisks.}
\label{fig:chiral fits}
\end{figure}

In Fig.~\ref{fig:GE and GM} we show  quenched and unquenched
 results for the isovector electric
and magnetic form factors using Wilson fermions at the three
values of the quark mass given in Table~\ref{table:parameters}.
Both quenched and  unquenched results decrease with the quark mass
yielding a larger slope at small $Q^2$, which is the expected behavior.
 However
both quenched and unquenched results for $G_E$ deviate more
from experiment than $G_M$ for the same quark mass.
 The two main uncertainties regarding  lattice results
are finite $a$-effects and the fact that the 
u- and d- quark masses are larger than  physical.
In the quenched theory one can envisage repeating the
calculation using a finer lattice in order to check
 whether finite $a$-effects can explain this behavior.
 Evaluation of  these form factors closer to the chiral limit
with Wilson fermions is problematic and
one will have to use other types of discretization for
the fermions  to study the dependence on lighter quark masses.
To be able to 
    directly compare with experiment, using the results of this work,
we need to extrapolate  $G_E$ and $G_M$ to the chiral limit.
The quark masses  employed in this work correspond to pion masses in the
range  560 to 410 MeV in the quenched theory and
690 to 380 MeV in the unquenched theory. Pion cloud effects are expected to be
small in this range of pion masses and therefore we expect 
a linear dependence of the results on  $m_\pi^2$. 
Our lattice data at these quark masses confirm this expectation
and therefore, to obtain results in the chiral limit, we
  extrapolate the form factors linearly in $m_\pi^2$.
We show the linearly extrapolated lattice results 
for $F_1$ and $F_2$ in Fig.~\ref{fig:chiral fits}.
We observe  agreement between  quenched and unquenched results
 at the chiral limit.
 In addition, there is good agreement
between  lattice results for $F_2$
and  experiment, with perhaps small deviations
at small $Q^2$. This is not the case for $F_1$
where the experimentally determined isovector $F_1$ decays faster
 as compared to the lattice results.
In a recent calculation, the quark mass
dependence of the isovector magnetic moment and
radii was determined. This was done
within a
chiral effective theory with explicit nucleon and $\Delta$ degrees of
freedom~\cite{QCDSF,chiral}. Therefore we can
extract the  relevant 
 low energy constants and counterterms that enter in the expressions
of these quantities in the effective chiral 
theory from fits to our form factors
and 
obtain results at the physical pion mass. Fitting to our lattice data
we obtain the curve  shown by the solid line
in Fig.~\ref{fig:chiral fits}. 
The details regarding the fits are given in Ref.~\cite{NN}. 
 The dashed lines give the maximal
 error band determined
by varying the values of the
fitted parameters by their errorbars.
The extrapolated value of the magnetic moment at the physical pion mass 
is in agreement with
experiment.  The resulting fits for the radii are also shown
in Fig.~\ref{fig:chiral fits}. The pion mass dependence of the  Dirac radius
is not well reproduced. Since this is related to the slope of $F_1$ this
is not surprising given that the lattice results have a different slope from
the experimental results and hardly show any quark mass dependence.

\section{V. N to $\Delta$ transition form factors}
The evaluation of the N-$\Delta$
matrix element, within the fixed sink approach,
 requires
 a new set of sequential propagators since in  the final state,
instead of the nucleon, we have the $\Delta$. However, once 
we produce the sequential propagators needed for the evaluation of
 the  electromagnetic N-$\Delta$ matrix element,
the axial one can be obtained with almost no additional 
computational cost using the same  sequential propagators.
The N to $\Delta$ transition 
involves no disconnected
diagrams and therefore lattice results can
be  directly compared to experiment.
 
\subsection{V.1. Electromagnetic transition form factors}
 To address the question
of possible deformation 
in the nucleon system the experiment of choice is electroproduction of
the $\Delta$ that measures the nucleon-$\Delta$
 transition amplitudes.
Non-zero quadrupole amplitudes are
thought to be connected with a  non-spherical nucleon or/and 
$\Delta$~\cite{cnp}.  
We can establish direct contact with experiment   by
 calculating the N to $\Delta$ transition form factors in lattice QCD.
To obtain accurate results that can provide a meaningful comparison
 to experiment
 two novel aspects are implemented:
 1) An optimal combination of three-point functions,
which allows momentum transfers
in a spatially symmetric manner
obtained by an appropriate choice of the
interpolating field for the
$\Delta$. This is similar
to the construction of the optimal nucleon source for the calculation
of the nucleon magnetic form factor, $G_M$, discussed in
the previous Section but more involved~\cite{PRL}.
 2) An overconstrained analysis using all
lattice momentum vectors contributing to a given $q^2$ value
in the extraction of the
three transition form factors analogous to what
was done for the elastic nucleon form factors.

The matrix element for the $\gamma^* N \> \rightarrow \> \Delta$
 transition with on-shell nucleon and $\Delta$ states and real or
virtual photons has the form~\cite{Jones73}
\be
 \langle \; \Delta (p',s') \; | j^\mu | \; N (p,s) \rangle =
 i   \sqrt{\frac{2}{3}} \biggl(\frac{m_{\Delta}\; m_N}{E_{\Delta}({\bf p}^\prime)\;E_N({\bf p})}\biggr)^{1/2}
  \bar{u}_\sigma (p',s') {\cal O}^{\sigma \mu} u(p,s) \;
\label{DjN}
\ee
where 
$ u_\sigma (p',s')$ is a spin-vector in the Rarita-Schwinger formalism.
$ {\cal O}^{\sigma \mu}$  can be decomposed in terms of the Sachs form factors
as
\be
{\cal O}^{\sigma \mu} =
  {\cal G}_{M1}(q^2) K^{\sigma \mu}_{M1}
+{\cal G}_{E2}(q^2) K^{\sigma \mu}_{E2}
+{\cal G}_{C2}(q^2) K^{\sigma \mu}_{C2} \;,
\ee
where the magnetic dipole, ${\cal G}_{M1}$, the electric quadrupole,
${\cal G}_{E2}$,
 and the Coulomb
quadrupole, ${\cal G}_{C2}$, form factors depend on the momentum
transfer $q^2 = (p'-p)^2$. The kinematical functions
$K^{\sigma \mu}$ in Euclidean space
are given in ref.~\cite{NDelta1}.
The ratios $R_{EM}$ or EMR and   $R_{SM}$ or CMR
in
 the  rest frame of the $\Delta$
are obtained from the Sachs form factors via
\be
 R_{EM}= -\frac{{\cal G}_{E2}(q^2)}{{\cal G}_{M1}(q^2)} \quad,
\hspace*{0.5cm}
 R_{SM}=-\frac{|{\bf q}|}{2m_\Delta}\;\frac{{\cal G}_{C2}(q^2)}{{\cal G}_{M1}(q^2)} \quad.
\label{CMR}
\ee

\begin{figure}[h]
\includegraphics[height=.27\textheight]{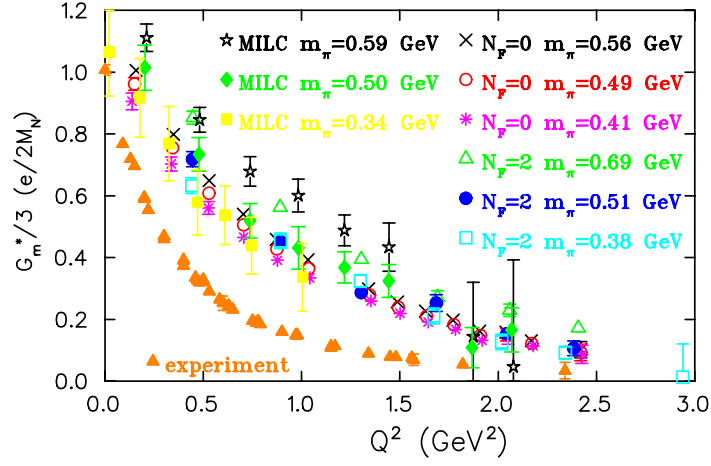}
\caption{$G_{m}^*(Q^2) \equiv
1/\sqrt{1+Q^2/(m_N+m_\Delta)^2}
 \>{\cal G}_{M1}(Q^2)$ as function of $Q^2$. 
The notation for Wilson fermions is the same 
as that in Fig.~\ref{fig:GE and GM}.
Results within the hybrid scheme
are shown with the stars
 for $m_l=0.03$, with
the filled rhombi for $m_l=0.02$ and with the filled squares for $m_l=0.01$.
Experimental results from Ref.~\cite{GM1-experiment} are shown by the filled triangles.}
\label{fig:GM1}
\end{figure}

\begin{figure}[h]
\begin{minipage}{3.5cm}
\hspace*{-5cm}
\includegraphics[height=.4\textheight,width=2.35\textwidth]{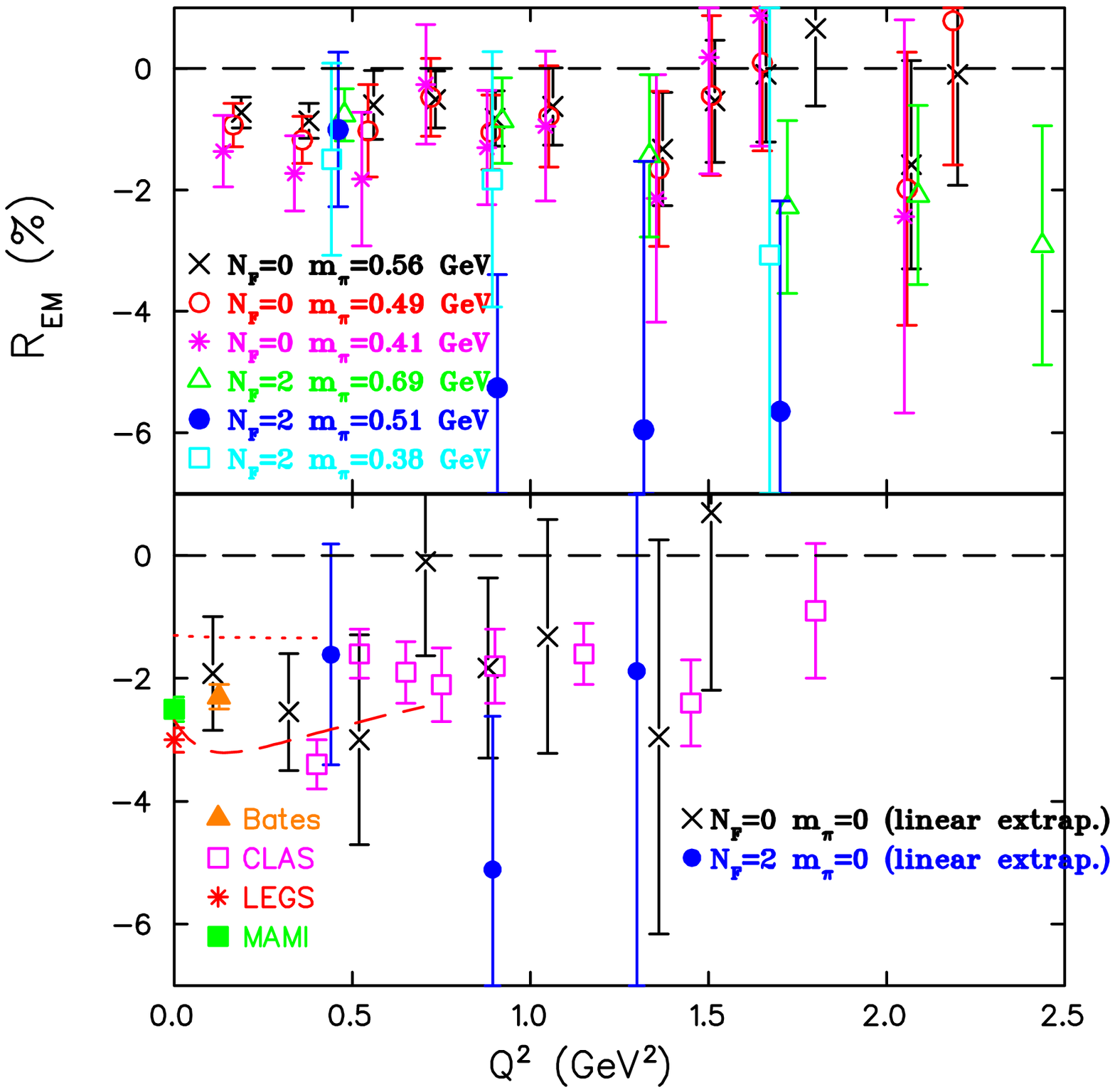}
\end{minipage}
\begin{minipage}{3.4cm}
\includegraphics[height=.4\textheight,width=2.35\textwidth]{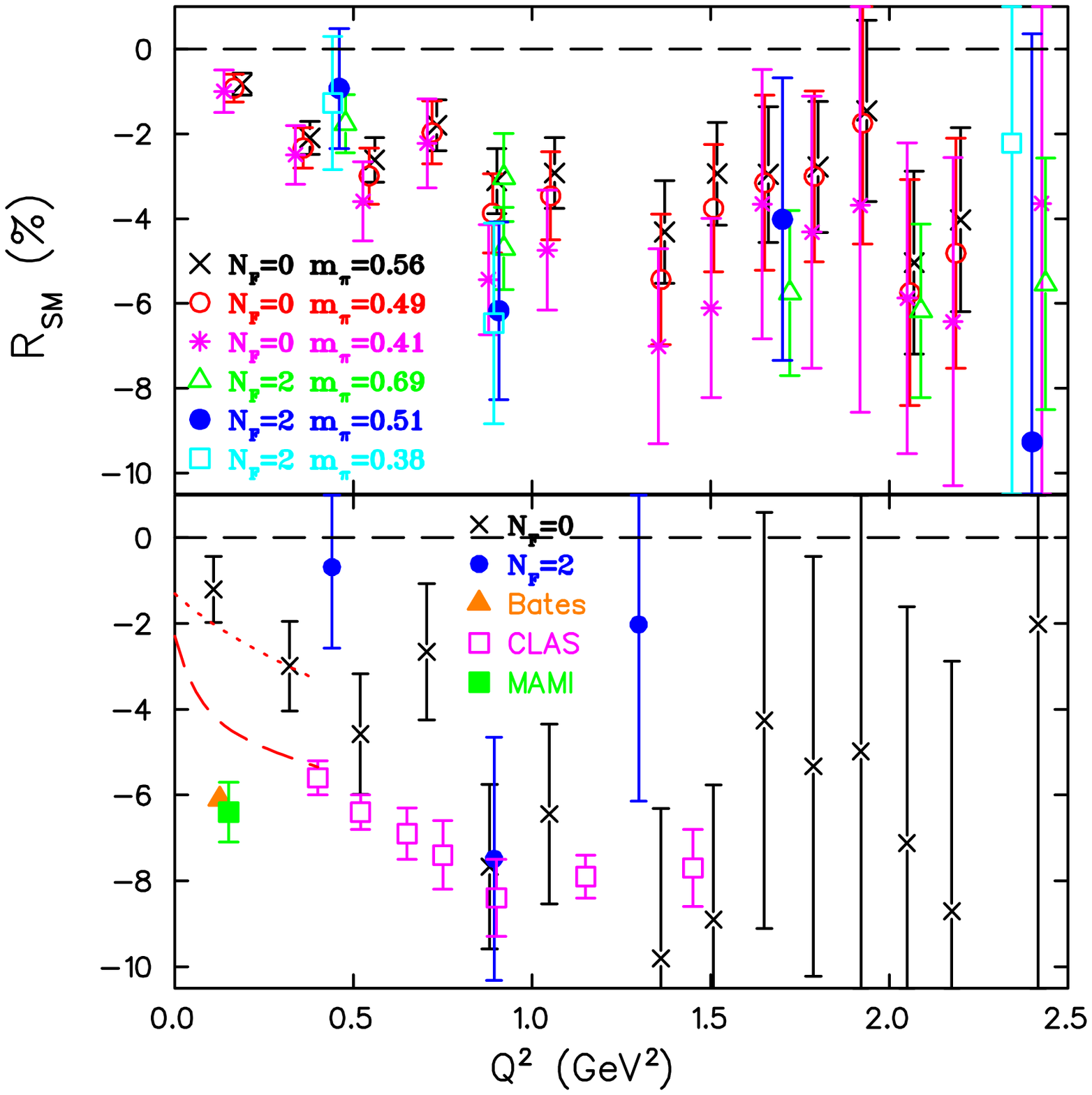}
\end{minipage}
\label{fig:EMR and CMR}
\caption{$R_{EM}$ (left) and $R_{SM}$ (right) as a function of $Q^2$.  
Upper graphs
show lattice results 
in the same notation as Fig.~\ref{fig:GE and GM}. The lower graphs show
 linear extrapolations to the chiral limit for quenched (crosses)
and unquenched (filled circles) Wilson fermions. 
We include recent experimental results from Refs.~\cite{Bates} (filled triangle), 
 \cite{Clas} (open squares), \cite{LEGS} (asterisk),
\cite{MAMI} (filled square). The  dotted and dashed lines are 
the results from a dynamical
model with bare and dressed vertices respectively~\cite{SL}.} 
\end{figure}

 We look for a plateau in
the large Euclidean
time behavior of the optimized ratio
\be
R_\sigma (t_2, t_1; {\bf q}\; ; \Gamma ; \mu)=
\frac{\langle G^{\Delta j^\mu N}_{\sigma} (t_2, t_1 ;  {\bf q};\Gamma ) \rangle \;}{\langle G^{\Delta }_{ii} (t_2, {\bf 0};\Gamma_4 ) \rangle \;} 
\biggl [ \frac{ \langle G^{N}(t_2-t_1, {\bf p};\Gamma_4 ) \rangle \;\langle
G^{\Delta }_{ii} (t_1, {\bf 0};\Gamma_4 ) \rangle \;\langle
G^{\Delta }_{ii} (t_2, {\bf 0};\Gamma_4 ) \rangle \;}
{\langle G^{\Delta}_{ii} (t_2-t_1, {\bf 0};\Gamma_4 ) \rangle \;\langle
G^{N} (t_1, {\bf p};\Gamma_4 ) \rangle \;\langle
G^{N} (t_2, {\bf p};\Gamma_4 ) \rangle \;} \biggr ]^{1/2} \quad.
\label{R-ratio2}
\ee
To extract ${\cal G}_{M1}$
we take the symmetric combination,
\be
S_1({\bf q};\mu)=  \sum_{\sigma=1}^3\Pi_\sigma({\bf q}\; ; \Gamma_4 ;\mu)
=i A\sum_{\sigma=1}^3  \epsilon^{\sigma 4\mu j} p^j {\cal
G}_{M1}(Q^2)
\label{GM1}
\ee
computed for all spatial current directions so that
lattice momentum vectors in all directions contribute. This combination,
is built into the $\Delta$ interpolating
field and  requires only one inversion.
Other combinations yield ${\cal G}_{E2}$ and 
$G_{C2}$~\cite{PRL,latt05_NDelta,tsapalis}.

In Fig.~\ref{fig:GM1} we show our results for  the magnetic dipole
form factor  at similar pion 
masses for  quenched and dynamical Wilson fermions
and in the hybrid approach 
using MILC configurations and domain wall fermions.
 Comparing results for quenched and unquenched Wilson fermions
we see that unquenching effects are small in this range of quark masses.
We also observe agreement between the results obtained 
in the two dynamical calculations.
As we have already pointed out, agreement between the results using dynamical
Wilson fermions and domain wall fermions is non-trivial since these
discretization schemes have different lattice systematics.
In all cases ${\cal G}_{M1}$ decreases with the quark mass
bringing lattice results closer to experiment. However, assuming a linear
dependence in  $m_\pi^2$ to extrapolate lattice data
 obtained using Wilson fermions where
 the statistical errors are the smallest, 
we find results at the chiral limit, which are higher than experiment.
Module finite $a$-effects, the conclusion is that  
 to reconcile lattice results with experiment a stronger 
dependence on $m_\pi^2$ for small quark masses
 seems necessary. In the hybrid approach the errors
are larger and a reduction in the errors by at least a factor of two is 
required
in order to draw any concrete conclusions regarding whether
they show a better agreement with 
experiment.

Results for the EMR and CMR ratios are only shown for Wilson fermions
in Fig.~\ref{fig:EMR and CMR}
since the errors on these ratios calculated in the hybrid scheme
are too large
prohibiting a meaningful comparison to the Wilson results. 
Although the
 results
from dynamical Wilson fermions have larger errors than in the quenched case
they tend to favor  negative non-zero values for both ratios. 
However, given the size of the errors,  an order of
magnitude more configurations need to be analyzed in order to
assess unquenching effects and  draw a definite  conclusion
about the values of these ratios  in the unquenched case. 
An analysis within
a chiral effective theory
leads to a non-trivial quark mass 
dependence for these ratios ~\cite{Marc} that brings  
quenched lattice data in agreement with experiment at the
lowest $Q^2$-value.

\subsection{V.2. Axial from factors}

The N to $\Delta$ transition, besides being used to probe
 electromagnetic properties of the nucleon system, is also well suited
 for studying the
weak structure functions.
This is because
 the $\Delta(1232)$, as the dominant nucleon resonance,
  can be well identified
and being a  purely isovector spin-flip transition, provides selective
information on hadron structure. 
 A lattice calculation of  
the axial form factors
is timely and important given that experiments at Jefferson Lab~\cite{Wells}
are underway to measure these form factors.
In this work we evaluate
the dominant contribution to  the parity violating 
asymmetry, which is 
determined by the ratio $C_5^A/C_3^V$ and is
to be measured at Jefferson Lab. This ratio
is the off-diagonal analog of the $g_A/g_V$ ratio 
extracted from neutron
$\beta$-decay and therefore  tests low-energy consequences of chiral symmetry, 
such as the
off-diagonal Goldberger-Treiman relation. 
In addition the ratio of axial 
form factors
$C_6^A/C_5^A$ provides
 a measure for the 
 conservation of the axial current~\cite{NDaxial}.

The invariant N to $\Delta$ weak matrix element can be expressed
in terms of four transition
form factors as~\cite{adler,LS}: 

\beq
<\Delta(p^{\prime},s^\prime)|A^3_{\mu}|N(p,s)> &=& i\sqrt{\frac{2}{3}} 
\left(\frac{M_\Delta M_N}{E_\Delta({\bf p}^\prime) E_N({\bf p})}\right)^{1/2}
\bar{u}^\lambda(p^\prime,s^\prime)\biggl[\left (\frac{C^A_3(q^2)}{M_N}\gamma^\nu + \frac{C^A_4(q^2)}{M^2_N}p{^{\prime \nu}}\right)  
\left(g_{\lambda\mu}g_{\rho\nu}-g_{\lambda\rho}g_{\mu\nu}\right)q^\rho \nonumber\\
&+&C^A_5(q^2) g_{\lambda\mu} +\frac{C^A_6(q^2)}{M^2_N} q_\lambda q_\mu \biggr]u(p,s)
\label{axial matrix element}
\eeq
where 
 $A^3_\mu(x)=  \bar{\psi}(x)\gamma_\mu \gamma_5 \frac{\tau^{3}}{2} \psi(x)$ 
is the isovector part of the axial current ($\tau^3$ being 
the third Pauli matrix). 
Having evaluated the electromagnetic N to $\Delta$ transition form factors
on the lattice  this matrix element can be evaluated
without requiring any further inversions 
since the optimized $\Delta$ sources are the same as those used in 
our study of the electromagnetic N to $\Delta$ transition 
and only  the operator
that couples to a quark line differs. 
 We consider the same ratio as that given in Eq.~\ref{R-ratio2} but 
replace the
three-point function $G_{\sigma}^{\Delta j_\mu N}(t,t_1;{\bf q};\Gamma)$ 
with $G_{\sigma}^{\Delta A^3_\mu N}(t,t_1;{\bf q};\Gamma)$. 
 In the large Euclidean time limit this ratio 
 yields the transition matrix element of 
Eq.~(\ref{axial matrix element}). 
Using, for example, the $\Delta$ source, $S_1$, for the evaluation
of the three-point function
 we obtain for large time separations $t_1$ and $t-t_1$  
\be
S_1({\bf q},\mu=4)=B\sum_{k=1}^3 p^k\Bigg[C^A_3+\frac{M_\Delta}{M_N}C^A_4
+\frac{E_N-M_\Delta}{M_N}C^A_6\Bigg]
\label{S1}
\ee
where the kinematical constants $B=B^\prime/\left[3(m_\Delta+m_N)\right]$
 and $B^\prime=\sqrt{2/3}
\left(M_\Delta/M_N + 1\right)\sqrt{\left(E_N({\bf p})+M_N\right)/E_N({\bf p})}$
and with kinematics where the  $\Delta$ is produced at rest.
Using $S_1$ and the other sink types used in our study of the
electromagnetic form factors~\cite{latt05_NDelta,NDaxial} 
the four axial form factors
 $C^A_3$, $C^A_4$, $C^A_5$ and $C^A_6$ can be determined
 by performing an
overconstrained analysis as described in Section IV.  

\begin{figure}[h]
\begin{minipage}{3.5cm}
\hspace*{-5cm}
\includegraphics[height=.43\textheight,width=2.35\textwidth]{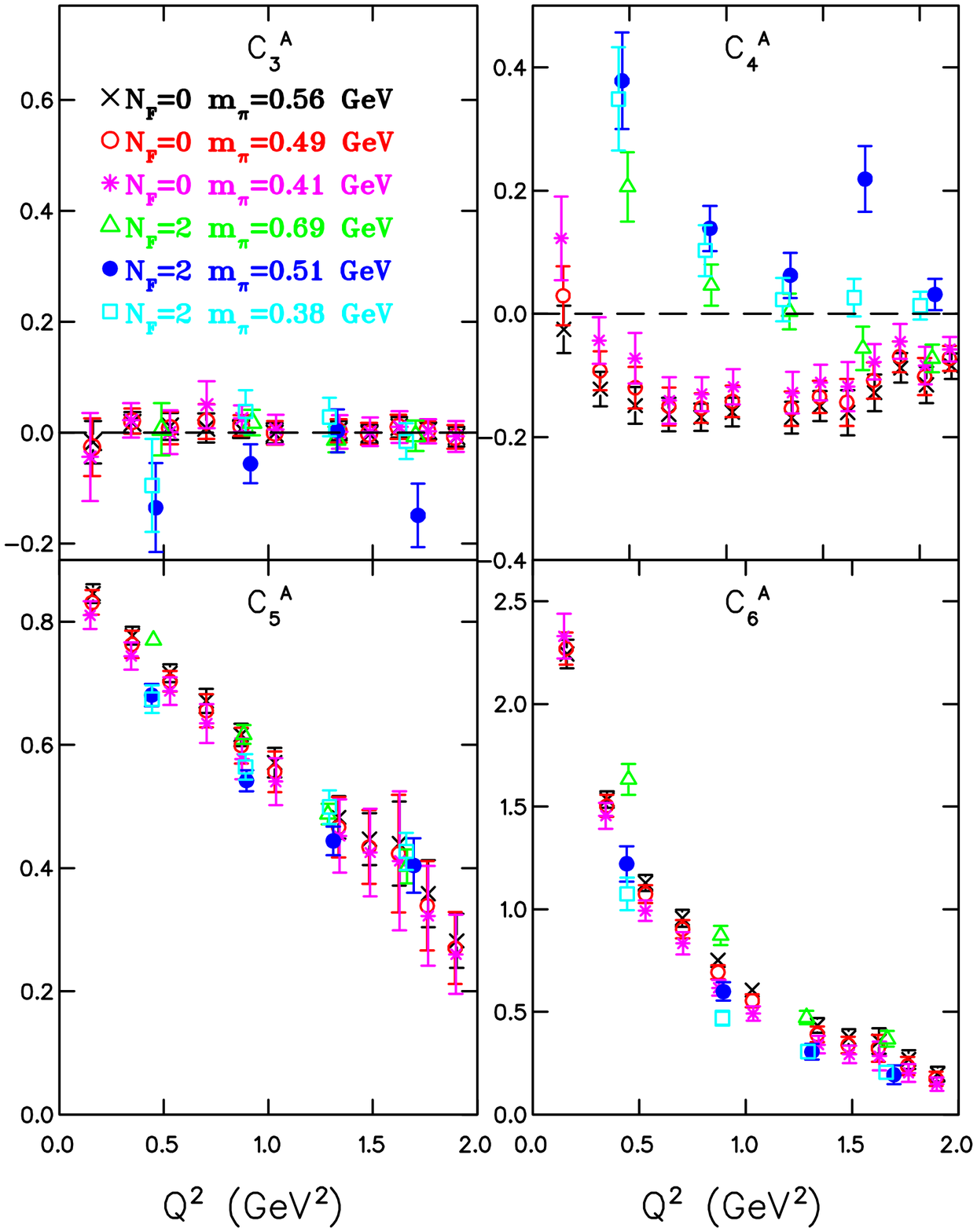}
\end{minipage}
\begin{minipage}{3.5cm}
\includegraphics[height=.43\textheight,width=2.35\textwidth]{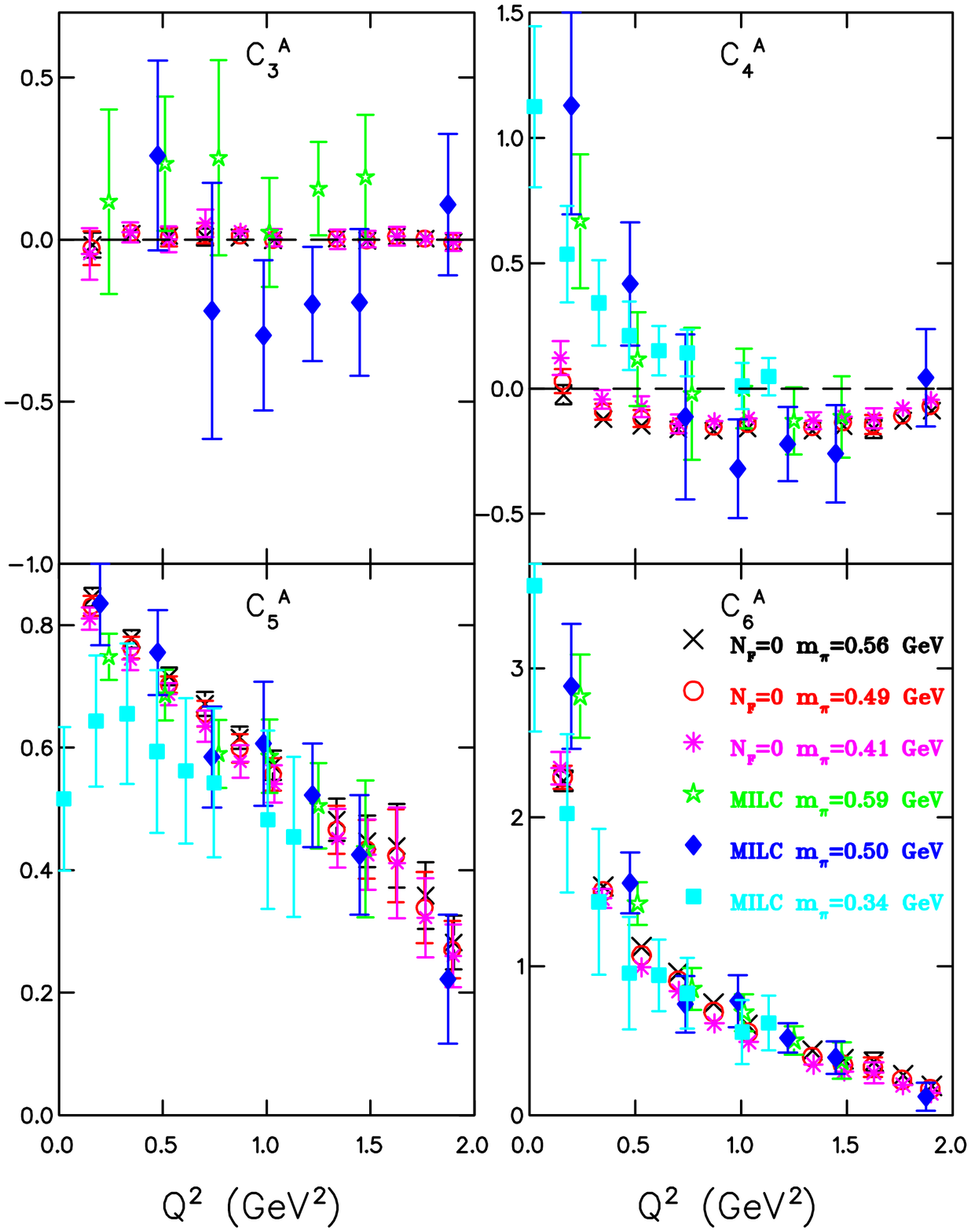}
\end{minipage}
\caption{The axial form factors $C^A_3$, $C^A_4$, $C^A_5$ and $C^A_6$
 as a function of $Q^2$.  
The notation is the same as that in Fig.~\ref{fig:GM1}.}
\label{fig:axial form factors}
\end{figure}

\begin{figure}[h]
\begin{minipage}{3.5cm}
\hspace*{-5cm}
\includegraphics[height=.3\textheight,width=2.35\textwidth]{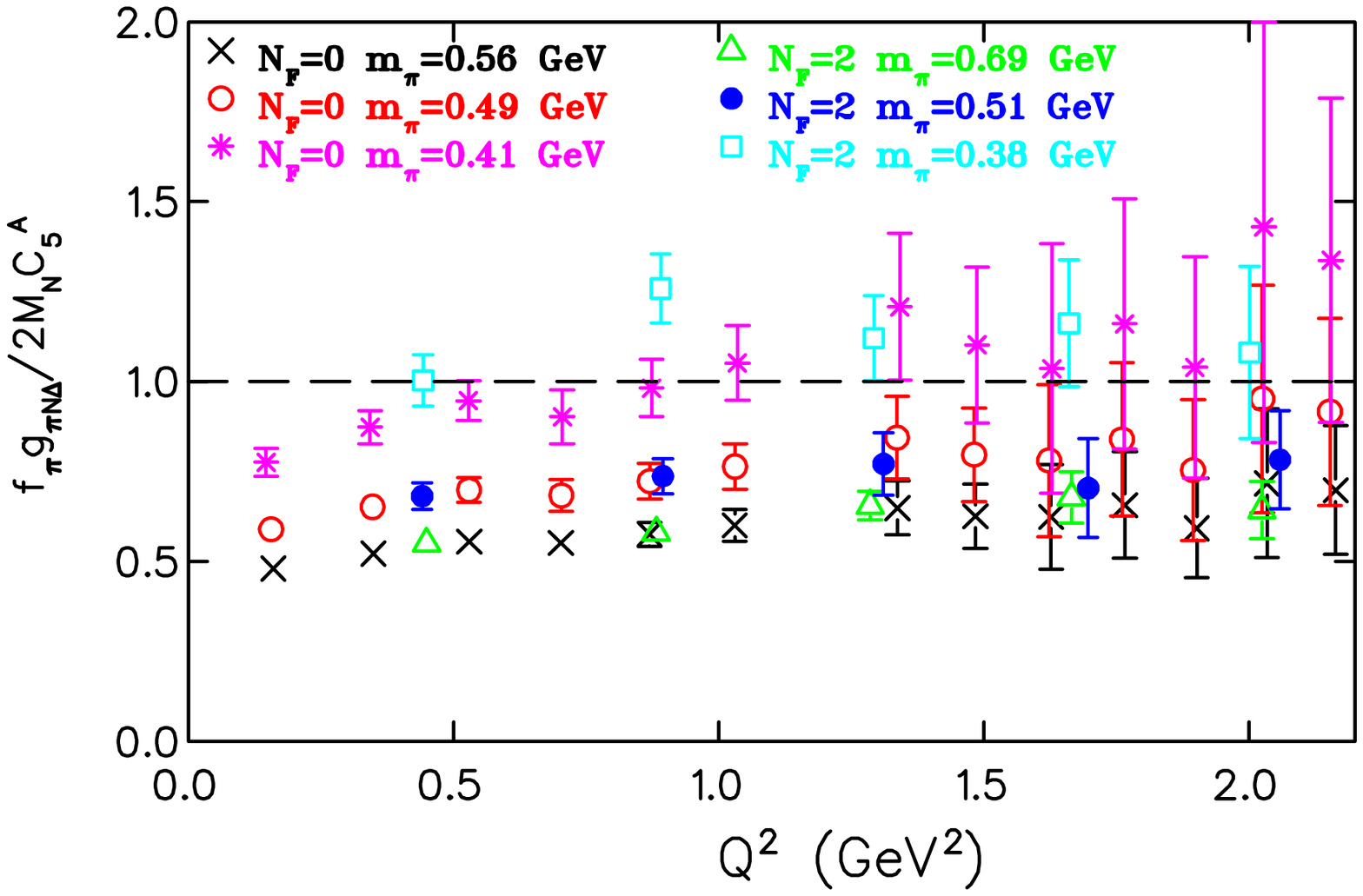}
\end{minipage}
\begin{minipage}{3.5cm}
\includegraphics[height=.3\textheight,width=2.35\textwidth]{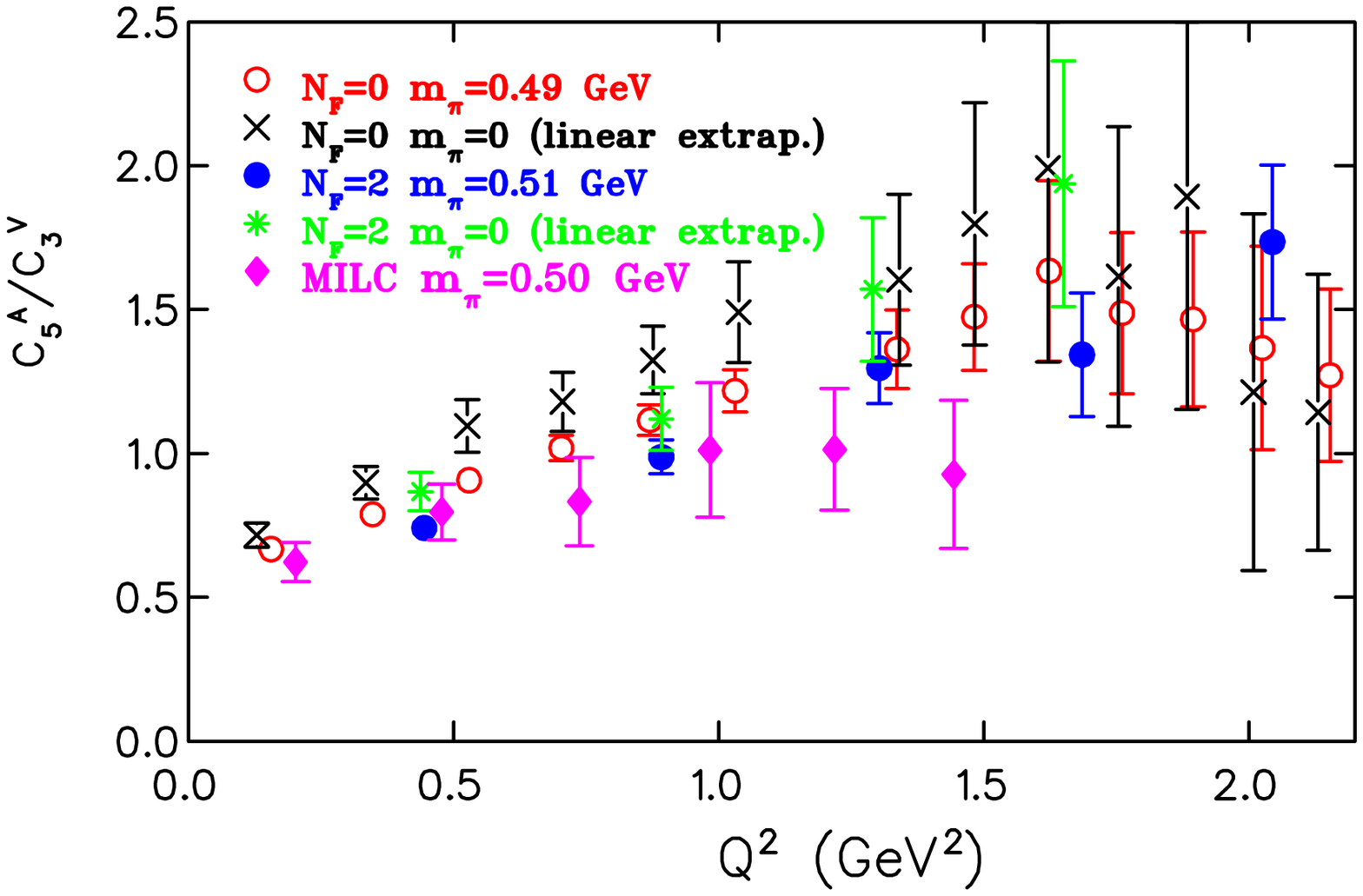}
\end{minipage}
\caption{
Left: The ratio $f_\pi g_{\pi N\Delta}/\left(2M_N C^A_5\right)$ is shown
 versus $Q^2$ for Wilson fermions in the same
notation as that of Fig.~\ref{fig:GE and GM}.
Right: The ratio $C^A_5/C^V_3$ is shown versus $Q^2$ 
 at $\kappa=0.1558$ (open circles) and 
at the chiral limit (crosses) for quenched QCD, 
at $\kappa=0.1580$ (filled circles)
and at the chiral limit (asterisks) for  
dynamical Wilson fermions  and 
at $am_l=0.02$ (filled rhombi) for domain wall fermions.}
\label{fig:chiral check}
\end{figure}

In Fig.~\ref{fig:axial form factors}  we 
show our lattice results for the four axial form factors for Wilson fermions
and in the hybrid scheme. 
For the Wilson fermions we use $Z_A=0.8$~\cite{renorm,renorm_dyn}.
In all cases  we observe that $C_3^A$ is 
consistent with zero. Comparison of quenched and unquenched
results obtained with Wilson fermions shows that unquenching effects 
are small for the dominant form factors, $C_5^A$ and $C_6^A$.
The form factor $C_4^A$ shows an interesting 
behavior: The unquenched results for both dynamical 
Wilson and domain wall fermions
show an increase at
low momentum transfers. Such large deviations between
quenched and full QCD results for these relatively heavy
 quark masses  are unusual
making  this  an interesting quantity to study 
effects of unquenching.

For finite mass pions partial conservation of axial current 
($\partial^\mu A_\mu^a(x)=f_\pi m_\pi^2\pi^a(x)$) leads to the
off-diagonal 
Goldberger-Treiman relation
$
C_5^A(Q^2)=f_\pi g_{\pi N\Delta}(Q^2)/2M_N
$
where $g_{\pi N\Delta}(Q^2)$ is determined
from the matrix element of the pseudoscalar density 
$<\Delta^+|\bar{\psi}(x)\gamma_5\frac{\tau^3}{2}\psi(x)|p>$
and the pion decay constant
$f_\pi$ from the two-point function $<0|A_4(x)|\pi>$. To relate the 
lattice pion matrix element to its physical value we need the
 pseudoscalar
renormalization constant, $Z_p$. For quenched~\cite{renorm} 
and dynamical Wilson fermions~\cite{renorm_dyn}
$Z_p(\mu^2a^2\sim 1)=0.5$, which however 
may depend on the renormalization scale.
At the chiral limit, at the lowest available momentum, we find 
$g_{\pi N\Delta}(Q^2=0.135\> {\rm GeV}^2)=18.0(1.9)$ and 
$g_{\pi N\Delta}(Q^2=0.443\> {\rm GeV}^2)=15.8(1.8)$ for  quenched and
dynamical Wilson fermions respectively where the quoted errors include a 10\%
uncertainty in $Z_p$. These values are 
to be compared with $g_{\pi N\Delta}(m_\pi^2)=23.2 \pm 2.6$ 
obtained from an analysis of $\pi N$ scattering~\cite{Meissner}.
 We show in Fig.~\ref{fig:chiral check} the ratio 
$f_\pi g_{\pi N\Delta}/\left(2 M_N C_5^A\right)$ for  Wilson fermions. 
We observe
that this ratio is almost independent  of $Q^2$  and
and that, as the quark mass decreases, it becomes consistent with
 unity in agreement with the off-diagonal Goldberger-Treiman relation. 

A prediction of lattice QCD is the ratio $C_5^A/C_3^V$. 
 The form factor $C_3^V$ can be obtained from the electromagnetic 
N to $\Delta$ transition. Using our lattice
results for the dipole and electric quadrupole Sachs factors, 
${\cal G}_{M1}$ and
${\cal G}_{E2}$, we extract
$C_3^V$ using the relation 
\be
C_3^V=\frac{3}{2}\frac{M_\Delta(M_N+M_\Delta)}{(M_N+M_\Delta)^2+Q^2}
\left({\cal G}_{M1}-{\cal G}_{E2}\right). 
\ee
The ratio $C_5^A/C_3^V$, 
shown in Fig.~\ref{fig:chiral check}  for $m_\pi\sim 500$~ MeV,
has  values  that fall within errorbars in the quenched theory and in full QCD.
Given this agreement between quenched and unquenching results in this
quark mass range, 
we opt to extrapolate the quenched results, which
carry the smallest errors, to the chiral limit. 
As can be seen from Fig.~\ref{fig:chiral check} a linear
extrapolation in $m_\pi^2$ leads to  only a small
increase in this ratio at the chiral limit. 
Without an analysis within a chiral effective theory for
the quark mass dependence,
 this is the best that can be done to extract 
 a reasonable estimate for the  physical value of this ratio.
 Under certain assumptions, such as 
taking $C_3^A\sim 0$ and considering that
$C_4^A$ is suppressed as
compared to $C_5^A$, both of which are justified
by the lattice results, the parity violating asymmetry can be shown to be
  proportional to 
this ratio~\cite{Nimai}. Our lattice results show  that this ratio and, to 
a first approximation the parity violating asymmetry, is non-zero
at $Q^2= 0$ and increases by a factor of 2-3 when $Q^2{\sim}1.5$~GeV$^2$.

\section{VI. Conclusions}
State-of-the-art lattice QCD calculations  
yield 
accurate results  
on a number of observables that are important for
  understanding the structure of the nucleon. 
We have presented the framework
of such lattice computations for the nucleon and N to $\Delta$ transition 
form factors for quenched and dynamical  Wilson fermions. The
results are accurate
enough to allow a meaningful comparison to experiment. 
We also evaluated the N to $\Delta$ transition form factors 
within a hybrid scheme, that combines 
the best simulation of the QCD vacuum that is available up to now using
staggered fermions, with domain wall fermions the have good chiral properties.
Within this hybrid scheme, 
we obtain   results which are consistent with those obtained
with dynamical Wilson fermions, albeit with larger statistical errors. 
Although chiral fermions are more expensive,
smaller quark masses can be reached  without any conceptual 
difficulties.
Work is in progress to calculate the $\Delta$ form factors and coupling
constants within the same framework.
 The aim of this program is to  evaluate 
a complete
set of observables for the nucleon-$\Delta$ system
at small enough
quark masses so that one can use  only lattice input
to fix the parameters of chiral effective theories. 
Extrapolation to the physical regime can then be carried out avoiding 
uncontrolled approximations.
 Dynamical Wilson and MILC configurations
at pion masses of 250 MeV on large enough volumes are now becoming 
available  and therefore, using the technology developed, we will be
able to calculate these quantities closer to the chiral limit where
we can reliably make contact with  chiral perturbation expansions.

\vspace*{-0.5cm}

\begin{theacknowledgments}
I am indebted to  my collaborators R. Edwards, 
Ph. de Forcrand, G. Koutsou, Th. Leontiou, 
Th. Lippert,  H. Neff, J. W. Negele,
K. Schilling, W. Schroers and A. Tsapalis
for their valuable input to this
work. I would like to thank C. N. Papanicolas for
bringing to my attention the 
experimental efforts to understand
hadron deformation and for providing
the motivation and encouragement to undertake the lattice QCD studies.
The computations were partly
carried out on the IBM machines at NIC, Julich,
Germany and at the National Energy Research Scientific
Computing Center (NERSC), 
 which is supported by the Office of Science of the U.S.
Department of Energy under Contract No. DE-AC03-76SF00098.
This work is
supported in part by the  EU Integrated Infrastructure Initiative
Hadron Physics (I3HP) under contract RII3-CT-2004-506078.

\end{theacknowledgments}



\bibliographystyle{aipprocl} 





\end{document}